\documentclass[12pt,a4paper]{article}
\pdfoutput=1

\usepackage[a4paper]{geometry}
\usepackage{amsmath}
\usepackage{amsfonts}
\usepackage{amssymb}
\usepackage{dsfont}
\usepackage{graphicx}
\usepackage[nosort]{cite}
\usepackage{slashed} 
\usepackage{mathtools}
\usepackage{multirow}
\usepackage[margin=10pt,font=small,labelfont=bf]{caption}

\allowdisplaybreaks[2]
\numberwithin{equation}{section}

\newcommand{\eq}[1]{\begin{equation}
                     \begin{split} #1 \end{split}
                     \end{equation}}

\newcommand{\op}{\hspace{1pt}}                     

\begin{document}


\thispagestyle{empty}

\begin{flushright}
  {\small
  LMU-ASC 75/17 \\
  MPP-2017-256
  }
\end{flushright}

\vspace*{0.75cm}


\begin{center}
{\LARGE
T-duality transformations for the  NS5-brane \\[8pt] along angular directions
}
\end{center}


\vspace{0.4cm}

\begin{center}
  Erik Plauschinn$^{\,1}$, Valent\'i Vall Camell$^{1,2}$
\end{center}


\vspace{0.4cm}

\begin{center} 
\emph{$^{1\,}$Arnold-Sommerfeld-Center for Theoretical Physics \\
Ludwig-Maximilians-Universit\"at M\"unchen \\ Theresienstra\ss e 37, 80333 M\"unchen \\ Germany} \\
\vspace{0.5cm}
\emph{$^{2\,}$Max-Planck-Institut f\"ur Physik \\
F\"ohringer Ring 6 \\ 80805 M\"unchen, Germany} \\
\end{center} 

\vspace{0.75cm}


\begin{abstract}
\noindent
In this note we study T-duality transformations for the NS5-brane
and its orbifolds along angular directions. We identify a geometric charge for these configurations and show
that it is interchanged with the $H$-flux under T-duality.
We furthermore perform a supersymmetry analysis and find that T-duality can break 
supersymmetry, in agreement with earlier results in the literature. 

\noindent
We  contrast our findings to compactifications of the NS5-brane on tori, 
which have vanishing geometric charge and for which T-duality transformations
along the compact directions preserve supersymmetry. 
This shows that the uncompactified NS5-brane and the compactified-and-smeared solution 
have different properties and behave differently under T-duality.

\end{abstract}

\clearpage


\tableofcontents


\section{Introduction}

One of the remarkable properties of string theory is its rich  structure of dualities: 
S-duality is a strong-coupling--weak-coupling duality, which for instance relates the type I  
to the heterotic $SO(32)$ string and under which type IIB string theory is self dual. 
T-duality relates circle-compactifications with radius $R$ to compactifications with radius
$\alpha'/R$, and establishes a connection between type IIA and type IIB string theory as well as
between the two heterotic theories. 
Furthermore, the type IIA and heterotic $E_8\times E_8$ theories arise from different 
compactifications of M-theory, and also the AdS/CFT duality plays an important role in
understanding 
string theory.

In this work we are interested in T-duality, which is reviewed for instance in \cite{Giveon:1994fu}. 
For $d$-dimensional toroidal string-theory backgrounds
with constant $B$-field the duality group is $O(d,d;\mathbb Z)$, and 
duality transformations can be performed explicitly by acting on the states
of the theory. For curved backgrounds the situation is more complicated, since in most cases a 
conformal-field-theory (CFT) description is not available.
However, 
T-duality transformation rules for curved backgrounds can be obtained by
following Buscher's procedure \cite{Buscher:1987sk,Buscher:1987qj,Buscher:1985kb} 
of identifying an isometry of the target-space geometry, gauging the corresponding 
symmetry in the world-sheet theory and integrating-out the gauge field.
When applying these Buscher rules to a background with 
non-vanishing $H$-flux, it turns out that   the dual 
geometry is topologically-different from the original background \cite{Bouwknegt:2003vb}. More concretely, 
if we specify the direction along which we dualize by a vector field $\mathsf{v}$, then under T-duality 
the two-form $\iota_\mathsf{v} H$ and the first Chern class of the circle-fibration $E_\mathsf{v}$ corresponding to $\mathsf{v}$ 
behave as
\eq{
  \iota_\mathsf{v} H \quad \xleftrightarrow{\hspace{10pt}\mbox{\scriptsize T-duality}\hspace{10pt}}
  \quad c_1(E_\mathsf{v}) \,.
}
Identifying then the non-triviality of the circle-fibration as a kind  of {\em geometric flux} \cite{Shelton:2005cf}, 
we see that 
under T-duality the $H$-flux and the geometric flux are interchanged.

In order to apply the Buscher rules a (compact) direction of isometry is needed. 
For a compact space $\mathcal M$ one naturally tries to identify a circle in $\mathcal M$
along which T-duality is performed, 
however, the formalism equally applies to angular isometries in a non-compact space. 
The latter setting has been briefly discussed for instance in \cite{Rocek:1991ps,Adams:2001sv} for the case of 
$\mathbb R^2$, where an angular T-duality transformation leads to a singular dual geometry.
It is expected that for such a geometry winding-modes of the string play an important role, but to 
our knowledge a complete understanding of this situation is still missing. 
It would therefore be worthwhile  to gain more insight
on angular T-duality transformations, which is something we want to address in this work.

More concretely, in this note we consider T-duality transformations along angular directions for the 
NS5-brane. Our motivation for studying this particular string-background is two-fold:
\begin{enumerate}

\item The NS5-brane solution is a well-known background with
non-trivial curvature and  $H$-flux.  The space transversal to the 
NS5-brane is four-dimensional and has an $\mathfrak{so}(4)$  isometry algebra.
However, different from the situation in flat space the norm of the Killing vectors 
does not vanish at the origin, and therefore an angular T-duality 
transformation of the NS5-brane does not (necessarily) lead to a singular dual space.
We can then compare this background to T-duals of flat space, in order to identify
the relevant degrees of freedom for understanding  the singularity of the latter.

\item T-duality transformations for the NS5-brane are usually performed by 
first compactifying  one or more of the directions  transversal to the NS5-brane solution, and in a second step 
smearing the localized NS5-brane  along the compact directions in order to 
obtain an  isometry (see for instance \cite{Gregory:1997te,Eyras:1998hn,Andreas:1998hh}). 
Compactifying the NS5-brane for instance on a circle, 
the isometry algebra gets broken from $\mathfrak{so}(4)$  to $\mathfrak{so}(3)$, 
while the smearing procedure enhances it to $\mathfrak{so}(3)\times \mathfrak{u}(1)$. 
Clearly, this changes the structure of the background.

The smeared NS5-brane solution compactified on a two-torus undergoes a similar 
change, reflected in the change of the isometry algebra $\mathfrak{so}(4)\to \mathfrak{so}(2)
\to\mathfrak{so}(2)\times \mathfrak{u}(1)\times \mathfrak{u}(1)$.
This smeared configuration is the starting point 
for a chain of T-duality transformations leading to the Kaluza-Klein (KK) monopole 
\cite{Gross:1983hb,Sorkin:1983ns} and 
the $5^2_2$-brane \cite{deBoer:2012ma} (also called $Q$-brane \cite{Hassler:2013wsa}),
\eq{
 \nonumber
 \setlength{\fboxsep}{3pt}
 \boxed{\begin{array}{c}\mbox{smeared NS5}\end{array}} 
 \hspace{10pt}\xleftrightarrow{\hspace{3pt}\mbox{\scriptsize T-duality}\hspace{3pt}}\hspace{10pt}
 \boxed{\begin{array}{c}\mbox{KK monopole}\end{array}} 
 \hspace{10pt}\xleftrightarrow{\hspace{3pt}\mbox{\scriptsize T-duality}\hspace{3pt}}\hspace{10pt}
 \boxed{\begin{array}{c}\mbox{$5^2_2$-brane}\end{array}} 
}
where the latter corresponds to a so-called non-geometric background 
\cite{Hellerman:2002ax,Dabholkar:2002sy,Hull:2004in,Shelton:2005cf,Lust:2015yia}. 
Corrections to the smearing-approximation for this chain of T-duality transformations
have been studied for instance in 
\cite{Gregory:1997te,Tong:2002rq,Harvey:2005ab,Witten:2009xu,Kimura:2013fda,Kimura:2013zva,Kimura:2014bea,Lust:2017jox}, and in the context of Double Field Theory (DFT) this point has been addressed 
for instance in \cite{Berman:2014jsa,Bakhmatov:2016kfn,Lust:2017jox,VallCamell:2017ogl}.
In this work we want to go one step further back and investigate T-duality transformations
for the uncompactified NS5-brane solution along angular directions. 
Ultimately our question is whether also here
a non-geometric background arises.

\end{enumerate}

\bigskip
This paper is organized in the following way: 
in order to prepare for the analysis of the NS5-brane, 
in section~\ref{sec_three_sphere} we discuss T-duality transformations for the 
three-sphere with $H$-flux along one and two directions. 
More concretely, in section~\ref{sec_s3_charges} 
we identify the NS-charge and we define geometric charges for this background.
In section~\ref{sec_s3_lens} we generalize 
these results by including orbifold projections,  in section~\ref{sec_s3_dual} 
we show how under T-duality the charges are interchanged
and in section~\ref{sec_s3_torus} we also determine the $\beta$-transform
of the three-sphere background.
Throughout our discussion we pay special attention to global issues.

In section~\ref{sec_ns5} we turn to the NS5-brane solution. 
In particular, in section~\ref{sec_ns_pre} we determine the NS-charge as well as the geometric
charges of the  NS5-brane and its orbifolds. 
In section~\ref{sec_ns_dual} we perform T-duality transformations along angular 
directions, and in section~\ref{sec_ns_susy} we analyze the amount of supersymmetry
preserved by these backgrounds. We find that one T-duality preserves at most one half of the 
original supersymmetry, while two T-dualities break supersymmetry completely.

Section~\ref{disc} contains a discussion of our results. First, we briefly compare our
findings with T-duality along angular directions for $\mathbb R^n$. 
After that, we contrast our results with the approach of compactifying the NS5-brane
solution, smearing along the compact directions and performing T-duality along the latter. 
Our general conclusion is that the uncompactified NS5-brane solution 
and the compactified-and-smeared solution have rather different properties and 
behave very differently under T-duality.

In three appendices we summarize some further details: in appendix~\ref{app_hopf} 
we discuss the global properties of the isometries of the three-sphere, 
in appendix~\ref{sec:Lens} we review lens spaces, and in appendix~\ref{sec:susy-app}
we give some details on our supersymmetry analysis of the NS5-brane solution and its
T-duals.


\section{T-duality for the three-sphere}
\label{sec_three_sphere}

For our discussion of T-duality transformations for the NS5-brane  it is
useful to review T-duality for the three-sphere with $H$-flux.
Some of the results presented in this section can already be found in the literature (see for instance \cite{Alvarez:1993qi,Gaberdiel:1995mx,Sarkissian:2008dq,Plauschinn:2014nha}), but in order 
to prepare for our subsequent analysis we want to recall them here.
We also carefully discuss some of the global issues related to these transformations, 
but the reader interested only in the NS5-brane and its T-duals may skip this section.


\subsection{The three-sphere}
\label{sec_s3_charges}

String theory on a three-sphere of radius $R^2=k$ with $k$ units of 
$H$-flux can be described by the $SU(2)$ Wess-Zumino-Witten (WZW) model at level $k$. 
Let us briefly review the main geometrical aspects of this background.

 
\subsubsection*{The setting}

We start by considering a round three-sphere of radius $R$, which  
can be defined by its embedding into $\mathbb{C}^2$ through the equation
\begin{equation}
|z_0|^2+|z_1|^2=R^2\,,
\end{equation}
where $(z_0,z_1)\in\mathbb C^2$.
Using Hopf coordinates
\begin{equation}\label{eq:hopf-coordinates}
z_0=R\op e^{i\xi_1}\cos\eta\,,
\hspace{60pt} 
z_1=R\op e^{i\xi_2}\sin\eta\,,
\end{equation}
with $\eta\in [0,\pi/2]$ and $\xi_1,\xi_2\in [0,2\pi)$, the metric and  $H$-flux of the $SU(2)_k$ WZW model 
take the following form
\eq{
\label{geo_001}
\arraycolsep2pt
\begin{array}{lcl}
ds^2&=& \displaystyle R^2 \op \bigl(\op d\eta^2+\cos^2\eta\,  d\xi_1^2+\sin^2\eta \, d\xi_2^2 \op\bigr) \,, \\[6pt]
H&=& \displaystyle 2\op k \sin\eta\,\cos\eta\, d\eta\wedge d\xi_1\wedge d\xi_2\,.
\end{array}
}
The dilaton is taken to be $\Phi=\phi_0={\rm const.}$
Note that  in order for the theory to be conformal one has to impose a relation between the radius and 
the level $k$ which reads
$R^2= |k|$.
However, for practical purposes it will be convenient for us to keep the dependence on $R^2$ 
explicit. 
In our subsequent analysis we will also make use of the following coordinate system with 
$\theta\in [0,\pi]$ and $\chi,\xi\in [0,2\pi)$
\begin{equation}\label{eq:modified-Hopf-coordinates}
\chi=\frac{1}{2}(\xi_1+\xi_2)\,,
\hspace{40pt} 
\xi=\xi_1-\xi_2\,,
\hspace{40pt}
\theta=2\eta\,.
\end{equation}
The metric and $H$-flux \eqref{geo_001} in these coordinates take the form
\eq{
\label{eq:modified-hopf-fields}
\arraycolsep2pt
\begin{array}{lcl}
ds^2&=& \displaystyle \frac{R^2}{4}\op \bigl(\op d\theta^2+4d\chi^2+d\xi^2-4\cos\theta \, d\chi \,d\xi\op\bigr )\,, \\[8pt]
H&=& \displaystyle \frac{k}{2}\op \sin\theta\, d\theta\wedge d\xi\wedge d\chi\,.
\end{array}
}


\subsubsection*{Isometries}

The isometry group of the round three-sphere is $O(4)$, and  therefore the isometry algebra is 
$\mathfrak{so}(4)\cong \mathfrak{su}(2)\times \mathfrak{su}(2)$. 
Note that this algebra  contains $\mathfrak{u}(1)\times \mathfrak{u}(1)$ as an abelian subalgebra. 
Using Hopf coordinates, 
the corresponding Killing vector-fields for these isometries are
\begin{equation}\label{eq:u1-vectors}
v=\partial_{\xi_1}+\partial_{\xi_2}=\partial_\chi\,,
\hspace{60pt}
\bar{v}=\partial_{\xi_1}-\partial_{\xi_2}=2\,\partial_\xi\,,
\end{equation}
which have a nowhere vanishing norm
\begin{equation}
|v|^2=|\bar{v}|^2=R^2 \,.
\end{equation}
These vector-fields can be integrated to $U(1)$ group-actions  on the three-sphere, 
and for $\lambda\in U(1)$ the group acts on the embedding coordinates $(z_0,z_1)$ 
in the following way
\eq{
\label{eq:u1 actions}
v:\hspace{15pt}&g_\lambda(z_0,z_1)=(\lambda \op z_0, \lambda \op z_1) \,,\\[4pt]
\bar{v}:\hspace{15pt}&\bar g_\lambda(z_0,z_1)=(\lambda \op z_0, \lambda^* z_1)\,.
} 
The orbits of such actions are $U(1)$ fibers everywhere in $S^3$, which we review in appendix~\ref{sec:Hopf}
in more detail.


\subsubsection*{Geometric charges}

The three-sphere can be described as a principal $U(1)$ bundle, where the fiber is along one of the directions in \eqref{eq:u1-vectors}. There is a natural procedure to assign a $U(1)$ connection to such fibrations \cite{Bossard:2008sw}.
In particular, for $U(1)$ fiber along some direction $\star$, we determine the corresponding connection as
\eq{
\mathcal{A}_{\star}=\frac{g_{\star i}}{g_{\star\star}}\,dx^i\,,
}
where $\{dx^i\}$ is a local basis on the co-tangent space. 
For the three-sphere we then obtain the following:
\begin{itemize}

\item Using the coordinates \eqref{eq:modified-Hopf-coordinates}, the metric 
in \eqref{eq:modified-hopf-fields} can be expressed as
\begin{equation}
\label{back_002}
ds^2=\frac{R^2}{4}\op \bigl(\op d\theta^2+\sin^2\theta\, d\xi^2\op\bigr)+R^2\left(d\chi-\frac{1}{2}\cos\theta \,d\xi\right)^2.
\end{equation}
The $U(1)$ gauge connection associated to the direction $\chi$ reads
\begin{equation}
\mathcal{A}_\chi=-\frac{1}{2}\cos\theta d\xi\,,
\end{equation}
and the corresponding field strength is computed as
\begin{equation}
\mathcal{F}_\chi=d\mathcal{A}_\chi=\frac{1}{2}\sin\theta\, d\theta\wedge d\xi \,.
\end{equation}
We can then define a geometric charge $n_{\chi}$ associated to the fibration  by integrating $\mathcal F_{\chi}$ 
over the base manifold $\mathcal B$. The latter is a two-sphere of radius $R/2$, and the charge is computed as
\begin{equation}\label{eq:geometric-charge-chi}
n_\chi=\frac{1}{2\pi}\int_{\mathcal B}\mathcal{F}_\chi  = 1\,.
\end{equation}
This is precisely the first Chern class of the fibration, which in general has to be an integer for principal 
$U(1)$ bundles.

\item Since also the direction $\xi$ in \eqref{eq:u1-vectors} corresponds to a $U(1)$ fiber, 
we can  compute the charge with respect to such a fibration structure. To do so, we note that the metric \eqref{eq:modified-hopf-fields} can be rewritten as
\begin{equation}\label{eq:hopf-metric-xi}
ds^2=\frac{R^2}{4}\op\bigl(\op d\theta^2+4\sin^2\theta\, d\chi^2\bigr)+\frac{R^2}{4}\op\bigl(\op d\xi-2\cos\theta \,d\chi\bigr)^2\,,
\end{equation}
from which we can determine the gauge field associated to this fiber-bundle structure as
\begin{equation}
\mathcal{A}_\xi=-2\cos\theta \,d\chi\,.
\end{equation}
This gauge field has field strength
\begin{equation}
\label{field_001}
\mathcal{F}_\xi=2\sin\theta\, d\theta\wedge d\chi\,,
\end{equation}
and the geometric charge we associate to this fibration has a different normalization as compared to \eqref{eq:geometric-charge-chi}. It is given by\footnote{
Using the definition \eqref{eq:geometric-charge-chi}, the charge associated with the field strength 
\eqref{field_001} would be $n=4$. 
However, the base-manifold $\mathcal B$ in \eqref{eq:hopf-metric-xi} is not a two-sphere 
but rather a double-cover. 
This has to be taken into account when computing the charge.}
\begin{equation}\label{eq:geometric-charge-xi}
n_\xi=\frac{1}{4\pi}\int_{\mathcal B}\mathcal{F_\xi} = 2\,.
\end{equation}
The fact that we obtain  two units of geometric charge is because there is an effective orbifolding in 
our coordinate $\xi$ (see appendix \ref{sec:Lens}). Due to this effect the charge along the coordinate $\xi$ in this coordinate frame is always even.

\end{itemize}


\subsubsection*{Remark}

Let us briefly note that in addition to the coordinates \eqref{eq:modified-Hopf-coordinates} there is one other
consistent choice, namely
\begin{equation}\label{eq:modified-Hopf-coordinates-tilde}
\tilde\chi=\xi_1+\xi_2\,,
\hspace{40pt} \tilde\xi=\frac{1}{2}(\xi_1-\xi_2)\,,
\hspace{40pt}
\tilde\theta=2\op\eta\,,
\end{equation}
again with periods $\tilde{\chi},\tilde{\xi}\in [0,2\pi)$ and $\tilde \theta\in[0,\pi]$.  This choice is one the same 
footing as \eqref{eq:modified-Hopf-coordinates},  and is more suitable to describe the background as a 
$U(1)_\xi$ fibration. The natural definitions for the geometric charges are now
\eq{
\tilde n_\chi=\frac{1}{4\pi}\int_{\mathcal B}\tilde{\mathcal{F}}_\chi\,,
\hspace{60pt}
\tilde n_\xi=\frac{1}{2\pi}\int_{\mathcal B}\tilde{\mathcal{F}}_\xi\,,
}
where $\tilde{\mathcal{F}}_\xi$ and $\tilde{\mathcal{F}}_\chi$ are the field strengths computed in the $(\tilde\chi,\tilde\xi)$ coordinates. In this  frame we obtain $\tilde{n}_\chi=2$ and $\tilde{n}_\xi=1$ for the three-sphere. 
However, the coordinates \eqref{eq:modified-Hopf-coordinates-tilde} will not 
play a major role in our subsequent discussion.


\subsubsection*{NS charge}

Apart from geometric charges, the background \eqref{eq:modified-hopf-fields} also has a non-trivial $H$-flux. Its associated charge is defined as  
usual as
\begin{equation}\label{eq:NS-charge}
h=\frac{1}{4\pi^2}\int_{S^3}H \,.
\end{equation}
In the case of the three-sphere \eqref{eq:modified-hopf-fields}, we find 
$h=k$ independent of the coordinate frame we choose. 
Furthermore, note that the $H$-flux is quantized as $h\in\mathbb{Z}$.


\subsection{Lens spaces and sphere orbifolds}\label{sec:orbifolds}
\label{sec_s3_lens}

A way to generalize the three-sphere background is 
to perform an orbifold projection.  
An orbifold constructed using a finite symmetry group of the space is again a manifold if and only if the action of any element of the group on the original space is a homeomorphism and the symmetry group acts freely on the space. These conditions are  satisfied if one consider a $\mathbb{Z}_{k}$ group acting on one of the $U(1)$ fibers of the three-sphere.


\subsubsection*{The $\mathbb{Z}_{k_1}^{(\chi)}$ orbifold}

We begin by considering the case where a discrete symmetry $\mathbb Z_{k_1}$ acts along the $U(1)_\chi$ fiber. The resulting space is locally the same as the original three-sphere, but is globally different. In particular, the orbifold background can be described by the same fields as in  \eqref{eq:modified-hopf-fields} but with the coordinate $\chi$ having the period
\begin{equation}
\chi\in \left[0,\frac{2\pi}{k_1} \right),
\hspace{60pt}
k_1\in \mathbb Z_+ \,.
\end{equation}
To restore the $2\pi$-periodicity we  rescale  $\chi\rightarrow k_1 \op \chi$, after which 
the resulting background fields read
\eq{
\label{eq:orbifold-chi}
\arraycolsep2pt
\begin{array}{lcl}
ds^2&=&\displaystyle \frac{R^2}{4}\left(d\theta^2+\frac{4}{k_1^2}\op d\chi^2+d\xi^2-\frac{4}{k_1}\op\cos\theta 
\,d\chi \,d\xi\right),\\[12pt]
H&=&\displaystyle \frac{k_3}{2}\op\sin\theta\, d\theta\wedge d\xi\wedge d\chi\,,
\end{array}
}
where now $\chi\in[0,2\pi)$ and where we defined $k_3=k/k_1$. In order for this model to be conformal, 
the radius has to satisfy $R=\sqrt{|k_1 k_3|}$. Such configuration is known as a lens space $L(k_1,1)$ (see appendix \ref{sec:Lens}) and corresponds to a $SU(2)_{k_1k_3}/\mathbb{Z}_{k_1}$ WZW model.
Next, we compute the  charges following the same procedure as in the previous section. With respect to the $U(1)_\chi$ fiber, the associated gauge field is
\begin{equation}
\mathcal{A}_\chi^{k_1}=-\frac{k_1}{2}\op\cos\theta \, d\xi\,,
\end{equation}
from which we determine the geometric charge as $n_\chi=k_1$.
The charge corresponding to the $H$-flux is determined as before and is given by
$h = k_3$.

Let us now observe that similarly as in \eqref{eq:modified-hopf-fields}, the orbifolded background
\eqref{eq:orbifold-chi} has in addition a  $U(1)_\xi$ isometry along the direction $\xi$. 
We can therefore try to determine the corresponding geometric charge. More concretely, 
for the connection we find 
\begin{equation}
\mathcal{A}_\xi^{k_1}=-\frac{2}{k_1}\cos\theta \,d\chi \,,
\end{equation}
whose charge, $n_\xi=2/k_1$, fails to be in $2\op\mathbb{Z}$. The reason is that after the orbifolding procedure, the $\xi$-fiber is not a $U(1)$ bundle anymore. Intuitively, the fiber along this directions remains a $U(1)$ everywhere except at the points $(z_0,z_1)=(0,1)$ and $(z_0,z_1)=(1,0)$, where the two fibers collide. There, the $\xi$-fiber becomes $U(1)/\mathbb{Z}_{k_1}$, and one cannot define a $U(1)$ connection 
along this direction.   


\subsubsection*{The $\mathbb{Z}_{k_2}^{(\xi)}$ orbifold}

A very similar discussion applies if one constructs an orbifold by acting with $\mathbb{Z}_{k_2}$ on the $U(1)_\xi$ fiber of the original $S^3$. In this case, the resulting configuration is 
\eq{\label{eq:orbifold-xi}
\arraycolsep2pt
\begin{array}{lcl}
ds^2&=&\displaystyle \frac{R^2}{4}\left(d\theta^2+4 d\chi^2+\frac{1}{k_2^2}\op d\xi^2-\frac{4}{k_2}\op\cos\theta \,d\chi\, d\xi\right),\\
H&=&\displaystyle \frac{k_3}{2}\op \sin\theta\, d\theta\wedge d\xi\wedge d\chi\,,
\end{array}
}
with $\chi,\xi\in[0,2\pi)$ and the radius has to satisfy $R^2=|k_2 k_3|$. One can now compute the charge with respect to the $\xi$ direction, for which the corresponding gauge field reads
\begin{equation}
\mathcal{A}_\xi^{k_2}=-2 \op k_2\cos\theta \,d\chi\,.
\end{equation}
Using the conventions \eqref{eq:geometric-charge-xi} one obtains for the charge $n_\xi=2k_2$, which satisfies $n_\xi\in 2\op\mathbb{Z}$. Analogous to what happened in the previous case, the configuration is not a $U(1)_\chi$ bundle anymore, therefore the charge is not well-defined.
Furthermore, as argued in appendix \ref{sec:Lens} the configuration \eqref{eq:orbifold-xi} is only well-defined for odd $k_2$. The reason is that, since the coordinate frame encodes an artificial orbifolding along the direction $\xi$, one can only describe further orbifoldings of this direction which are compatible with it. This is the case for odd $k_2$. If one wants to construct $\mathbb{Z}_{k_2}^{(\xi)}$ orbifolds with even $k_2$, one needs to use the frame \eqref{eq:modified-Hopf-coordinates-tilde}, where the configuration would read
\eq{
\arraycolsep2pt
\begin{array}{lcl}
ds^2&=&\displaystyle \frac{R^2}{4}\left(d\theta^2+ d\tilde\chi^2+\frac{4}{k_2^2}\op d\tilde\xi^2-\frac{4}{k_2}\op\cos\theta \,d\tilde\chi \,d\tilde\xi\right),\\[10pt]
H&=&\displaystyle\frac{k_3}{2}\sin\theta\,\, d\theta\wedge d\tilde\xi\wedge d\tilde\chi\,.
\end{array}
}


\subsubsection*{The $\mathbb{Z}_{k_1}^{(\chi)}\times\mathbb{Z}_{k_2}^{(\xi)}$ orbifold}

Finally, let us comment on the possibility of having $\mathbb{Z}_{k_1}\times\mathbb{Z}_{k_2}$ orbifolds. Since the $U(1)\times U(1)$ isometries collapse into a single $U(1)$ at two points on the three-sphere, one has to worry whether it is possible that $\mathbb{Z}_{k_1}\times\mathbb{Z}_{k_2}$ acts freely on these points. In fact, this is the case if $k_1$ and $k_2$ are relatively prime, and the group will act as a $\mathbb{Z}_{k_1k_2}$ (see appendix \ref{sec:Lens}). However, since the group acts as $\mathbb{Z}_{k_1k_2}$ on colliding $U(1)$'s but as $\mathbb{Z}_{k_1}\times\mathbb{Z}_{k_2}$ elsewhere,  none of the directions $\chi$ or $\xi$ will be global isometries.

Following the discussions above, we write the most general orbifold configuration using the coordinates \eqref{eq:modified-Hopf-coordinates} as
\eq{\label{eq:general-orbifold}
\arraycolsep2pt
\begin{array}{lcl}
ds^2&=&\displaystyle\frac{R^2}{4}\left(d\theta^2+\frac{4}{k_1^2}\op d\chi^2+\frac{1}{k_2^2}\op d\xi^2-\frac{4}{k_1k_2}\op \cos\theta \,d\chi \,d\xi\right),\\[10pt]
H&=&\displaystyle \frac{k_3}{2}\op \sin\theta\, d\theta\wedge d\xi\wedge d\chi \,,
\end{array}
}
or using the frame \eqref{eq:modified-Hopf-coordinates-tilde} as
\eq{\label{eq:general-orbifold-1}
\arraycolsep2pt
\begin{array}{lcl}
ds^2&=&\displaystyle\frac{R^2}{4}\left(d\theta^2+\frac{1}{k_1^2}\op d\tilde\chi^2+\frac{4}{k_2^2}\op d\tilde\xi^2-\frac{4}{k_1k_2}\op \cos\theta \,d\tilde\chi \,d\tilde\xi\right),\\[10pt]
H&=&\displaystyle \frac{k_3}{2}\op\sin\theta\, d\theta\wedge d\tilde\xi\wedge d\tilde\chi\,.
\end{array}
}
Note that the first configuration fails to describe the cases with even $k_2$ and the second the cases with even $k_1$. In both situations the radius is $R=\sqrt{|k_1k_2k_3|}$. A direct computation shows that the geometric charges are $n_\chi=\frac{k_1}{k_2}$ and $n_\xi=\frac{2 k_2}{k_1}$ for the first case and $\tilde n_\chi=\frac{2k_1}{k_2}$ and $\tilde n_\xi=\frac{k_2}{k_1}$ for the second. Except for  $k_1=1$ or $k_2=1$, all of them are non-integers since $k_1$ and $k_2$ are relatively prime.


\subsection{Clifford tori}\label{sec:tori}
\label{sec_s3_torus}

As it can be inferred  for instance from the metric in \eqref{geo_001}, the three-sphere can locally be seen as a two-torus -- constructed with the two $U(1)$ directions -- fibered over a line segment. This is not a globally-defined fibration structure, since there are two points on the base where the torus degenerates to a circle. 
However,  this picture is nevertheless useful since we can interpret T-duality transformations  
as $O(2,2;\mathbb Z)$ acting on the $\mathbb T^2$.


\subsubsection*{K\"ahler and complex structure of $\mathbb T^2$}

Let us therefore parametrize the two-torus  in terms of its complex and K\"ahler structure $\tau$ and $\rho$ as
\begin{equation}\label{eq:tau-rho-definitions}
\tau=\frac{g_{\chi\xi}}{g_{\chi\chi}}+i\frac{\sqrt{\det g_{\mathbb T^2}}}{g_{\chi\chi}}\,,\hspace{60pt}\rho=B_{\mathbb T^2}+i\sqrt{\det g_{\mathbb T^2}}\,.
\end{equation}
In terms of these parameters the general configuration \eqref{eq:general-orbifold} is described by
\begin{equation}\label{eq:tau-rho-3sphere}
\tau=-\frac{1}{2}\, \frac{k_1}{ k_2}\op e^{-i\theta}\,,\hspace{60pt}\rho=-\frac{1}{2}\,k_3\op e^{-i\theta}\,,
\end{equation}
where $\theta\in [0,\pi]$ is the coordinate along the line segment. Note that at the end-points of the segment the imaginary parts of $\tau$ and $\rho$ vanish, which are the points where one of the two cycles of the torus collapses. For the following analysis, we will also consider the coordinate system \eqref{eq:modified-Hopf-coordinates-tilde}, in which the parameters \eqref{eq:tau-rho-definitions} for \eqref{eq:general-orbifold-1} are
\begin{equation}\label{eq:tau-rho-tilde-3sphere}
\tilde\tau=-2\,\frac{k_1}{k_2}\op e^{-i\theta}=4\op\tau\,,\hspace{60pt}\tilde\rho=-\frac{1}{2}\,k_3\op e^{-i\theta}=\rho\,.
\end{equation}
Finally, for all three-spheres the  component of the metric on the base is required to be $R^2=|k_1k_2k_3|$. This is not affected by any transformation of the toroidal coordinates, and hence a transformation that preserves $R^2$  will preserve the three-sphere structure.


\subsubsection*{$SL(2,\mathbb{Z})_\tau$ transformations}

To get some understanding of the three-sphere from the point of view of a 
torus fibration, we analyze how $SL(2,\mathbb{Z})$ transformations act on the complex structure $\tau$. These transformations are large diffeomorphisms, and therefore we can also understand them as transformations acting on the coordinates $(\xi,\chi)$. However, as already discussed, this coordinate frame has some orbifold structure intrinsically encoded in it, which also needs to be taken into account.

In particular, let us analyze the transformation of the form $\tau\rightarrow-1/\tau$, which corresponds to a $\pi/2$-rotation of the coordinates. To do this, we will rely on the original Hopf coordinates \eqref{eq:hopf-coordinates}, where the two angular directions form a torus with no additional structure. 
Under this transformation, the coordinates $(\xi,\chi)$ transform as
\begin{equation}\label{eq:chi-xi-rotation}
\xi\rightarrow -2\chi=-\tilde{\chi}\,,\hspace{60pt}\chi\rightarrow \frac{1}{2}\xi=\tilde{\xi}\,,
\end{equation}
and the orbifolding structure is also rotated. Therefore, the natural transformation to study is
\begin{equation}
\tau\rightarrow-\frac{1}{\tilde{\tau}}=-\frac{1}{4\op\tau},
\end{equation}
which is not an $SL(2,\mathbb{Z})$ transformation anymore, but is indeed the transformation obtained by conjugating  $\tau\rightarrow -1/\tau$ with the coordinate transformation \eqref{eq:modified-Hopf-coordinates}. 
Applying this transformation to \eqref{eq:tau-rho-3sphere}, the resulting configuration is described by
\begin{equation}
\tau '=\frac{1}{2}\, \frac{k_2}{ k_1}\op e^{i\theta}\,,
\end{equation}
which acts on the integer numbers $k_1$ and $k_2$ by
\begin{equation}\label{eq:tau-k-transformations}
k_1'=k_2\,,\hspace{60pt} k_2'=-k_1\,,
\end{equation}
as one would have naturally guessed. The correct sign for $\text{Im}\,\tau '$ is obtained by taking absolute values of $k_i'$. This transformation preserves the radius $R$ of the three-sphere and, therefore, the resulting configuration is again a three-sphere orbifold.

Finally, we point-out that the same result can also be obtained by an explicit rotation of the metric, after which the geometric charges in the conventions \eqref{eq:geometric-charge-chi} and \eqref{eq:geometric-charge-xi} are
$n_\chi=-\frac{k_2}{k_1}$ and $n_\xi=-\frac{2 k_1}{k_2}$, 
which is in agreement with  \eqref{eq:tau-k-transformations}.


\subsubsection*{$\beta$-transformation}

Since we have a description of the three-sphere in terms of a torus fibration, we 
want to take the opportunity to perform a $\beta$-transformation of the background.
This is related to the $\beta$-deformation first discussed in \cite{Lunin:2005jy}.
In terms of  the modular parameters of the two-torus, such a transformation is given by 
\begin{equation}
\rho\longrightarrow\frac{\rho}{-\beta\op\rho+1} \,,
\end{equation}
where $\beta\in\mathbb{Z}$. Since the parameter $\rho$ is not affected when changing between the different frames, the resulting configuration can be obtained in all of them in an analogous way. In particular, applying this transformation to \eqref{eq:general-orbifold} leads to the configuration
\eq{
\label{beta_sphere_1}
\arraycolsep2pt
\begin{array}{lcl}
ds^2&=&\displaystyle\frac{R^2}{4}d\theta^2+\frac{R^2}{\Lambda}\left(\frac{4}{k_1^2}\op d\chi^2+\frac{1}{k_2^2}\op d\xi^2-\frac{4}{k_1k_2}\op \cos\theta \,d\chi \,d\xi\right),\\[14pt]
H&=&\displaystyle \frac{2\op(\beta ^2 k_3^2-4)\op k_3 \sin\theta}{\Lambda^2}\op \, d\theta\wedge d\chi\wedge d\xi \,,\\[15pt]
e^{2\Phi}&=& \displaystyle e^{2\phi_0}\,\frac{4}{\Lambda}\,,
\end{array}
}
where $\phi_0$ is the constant dilaton of the original three-sphere orbifold and
where we defined
\eq{
\label{beta_sphere_2}
  \Lambda = 4+\beta ^2\op k_3^2+4 \op\beta\op k_3 \cos\theta \,.
}
The  background \eqref{beta_sphere_1} is a solution of the supergravity equations of motion,
and is non-singular except for $|\beta\op k_3|=2$.


\subsection{T-duality}
\label{sec_s3_dual}

We now want to investigate the configurations obtained after applying a factorized T-duality on the sphere and its orbifolds. Buscher's approach to T-duality involves gauging isometries of the background \cite{Buscher:1987sk,Buscher:1987qj,Buscher:1985kb}, and we 
therefore expect to find globally well-defined T-dual spaces when the isometry of the original background 
is globally defined.


\subsubsection*{T-duality along $\chi$: $\tau\leftrightarrow\rho$}
We start by considering T-duality transformations along the direction $\chi$, which in the conventions \eqref{eq:tau-rho-definitions} with coordinate frame \eqref{eq:modified-Hopf-coordinates} corresponds to the interchange $\tau\leftrightarrow\rho$. The configuration dual to \eqref{eq:tau-rho-3sphere} is characterized by the parameters
\begin{equation}
\tau '=-\frac{1}{2}\, k_3\op e^{-i\theta},\hspace{60pt}\rho '=-\frac{1}{2}\, \frac{k_1}{k_2}e^{-i\theta}\,, 
\end{equation} 
with corresponding charges
\begin{equation}
n_\chi '=k_3\,,\hspace{50pt}n_\xi '=\frac{2}{k_3}\,,\hspace{50pt} h'=\frac{k_1}{k_2}\,.
\end{equation}
The condition $h'\in\mathbb{Z}$ can only be satisfied if $k_2=1$, 
since
the original background is globally well-defined if and only if $k_1$ and $k_2$ are coprime.
In fact, as discussed in section \ref{sec:orbifolds}, this is the only case when the isometry along $\chi$ of the original configuration is globally-defined. Therefore, dualizing along it when $k_2\neq 1$ leads to a background globally ill-defined. If the condition $k_2=1$ is satisfied, the radius $R$ of the three-sphere remains invariant and the resulting configuration is a three-sphere orbifold with charges $(n_\chi ',h')=(h,n_\chi)$ \cite{Sarkissian:2008dq}.

The same results can be also obtained by direct application of Buscher rules to the configuration \eqref{eq:general-orbifold} with $k_2=1$, obtaining
\eq{
\arraycolsep2pt
\begin{array}{lcl}
ds^2&=&\displaystyle\frac{k_1 k_3}{4}\left(d\theta^2+\frac{4}{k_3^2}\op d\chi^2+d\xi^2-\frac{4}{k_3}\op\cos\theta \,d\chi\op d\xi\right),\\[12pt]
H&=&\displaystyle\frac{k_1}{2}\op\sin\theta\, d\theta\wedge d\xi\wedge d\chi \,,
\end{array}
}
and we can then conclude that the effect of T-duality along the direction $\chi$ is the interchange $k_1\leftrightarrow k_3$, provided $k_2=1$. Therefore, in this case, T-duality relates the following two conformal theories \cite{Maldacena:2001ky}
\begin{equation}
\frac{SU(2)_{k_1k_3}}{\mathbb{Z}_{k_1}}\quad\longleftrightarrow\quad\frac{SU(2)_{k_1k_3}}{\mathbb{Z}_{k_3}}\,.
\end{equation}


\subsubsection*{T-duality along $\tilde\xi$: $\tilde\tau\leftrightarrow-1/\tilde\rho$}
Next, we consider the T-duality transformation along direction $\xi$. The natural coordinate-frame to describe this duality is \eqref{eq:modified-Hopf-coordinates-tilde}, where it corresponds to the interchange $\tilde\tau\leftrightarrow -1/\tilde\rho$ in \eqref{eq:tau-rho-tilde-3sphere}. The dual background is then characterized by
\begin{equation}
\tilde\tau '=2\,\frac{1}{k_3}e^{i\theta}\,,\hspace{60pt}\tilde\rho '=\frac{1}{2}\, \frac{k_2}{k_1}e^{i\theta}\,, 
\end{equation}
with charges
\begin{equation}
\tilde n_\chi '=-\frac{2}{k_3}\,,\hspace{50pt}\tilde n_\xi '=-k_3\,,\hspace{50pt} \tilde h'=-\frac{k_2}{k_1}\,,
\end{equation}
and again the condition $\tilde{h}\in\mathbb{Z}$ is only satisfied when $k_1=1$, which is the case where the isometry in the original configuration is globally well-defined.

As in the case of T-duality along $\chi$, the same results can be obtained by direct application of Buscher rules to the configuration \eqref{eq:general-orbifold-1}. More interestingly, it is possible to obtain an equivalent result within the coordinate-frame of \eqref{eq:general-orbifold}. In this frame, the Killing vector 
is $\bar v =2\partial_\xi$ (see \eqref{eq:u1-vectors}), 
and applying the Buscher rules gives the T-dual background\footnote{For T-duality transformations 
along Killing vector-fields which are  not normalized to one, see for instance \cite{Plauschinn:2013wta,Plauschinn:2014nha}.}
\eq{
\arraycolsep2pt
\begin{array}{lcl}
ds^2&=&\displaystyle\frac{k_2 k_3}{4}\left(d\theta^2+4\op d\chi^2+\frac{1}{k_3^2}\op d\xi^2+\frac{4}{k_3}\op\cos\theta\, d\xi\op d\chi\right), \\[12pt]
H&=&\displaystyle -\frac{k_2}{2}\op\sin\theta\, d\theta\wedge d\xi\wedge d\chi\,,
\end{array}   
}
and one can use the previously-mentioned procedure to compute the charges to obtain
\begin{equation}
n'_\xi=-2\op k_3\,,\hspace{50pt} h'=-k_2\,,
\end{equation}
which are consistent with the results found using the other coordinate-frame. Furthermore, these results could  have been obtained also by considering the transformation $\tau\leftrightarrow-1/4\rho$. We then conclude that the effect of T-duality along the direction $\xi$, with $k_1=1$, is the interchange $k_2\leftrightarrow k_3$ and relates the conformal theories
\begin{equation}
\frac{SU(2)_{k_2k_3}}{\mathbb{Z}_{k_2}}\quad\longleftrightarrow\quad\frac{SU(2)_{k_2k_3}}{\mathbb{Z}_{k_3}}\,.
\end{equation}


\subsubsection*{T-duality along $\chi$ and $\xi$: $\tau\leftrightarrow-1/4\tau$, $\rho\leftrightarrow-1/4\rho$}

Finally, we consider factorized dualities along both directions of the torus simultaneously. Consecutively applying T-duality along the directions $\chi$ and $\xi$ corresponds to the transformation
(in terms of the coordinate frame\eqref{eq:modified-Hopf-coordinates})
\begin{equation}
\tau\rightarrow- \frac{1}{4\op\tau}\,,\hspace{50pt} \rho\rightarrow -\frac{1}{4\op\rho}\,.
\end{equation}
Applying these transformations to \eqref{eq:general-orbifold} one obtains a configuration where the fibered torus is described by
\begin{equation}
\tau'=\frac{1}{2}\,\frac{k_2}{k_1}e^{i\theta}\,,\hspace{50pt}\rho'=\frac{1}{2}\, \frac{1}{k_3}e^{i\theta}\,,
\end{equation}
and corresponding charges read
\begin{equation}
n'_\chi=-\frac{k_2}{k_1},\hspace{50pt}n'_\xi=-\frac{2k_1}{k_2},\hspace{50pt}h'=-\frac{1}{k_3}.
\end{equation}
The NS charge is  properly quantized only for $k_3=1$. This is in fact the self-dual point for the $\rho$-transformation and also the case where the radius $R$ of the three-sphere remains invariant. We emphasize that it is enough to have only one of the geometric charges correctly quantized in order to have at least one integer geometric charge in the dual background. In fact, the dual background for the case $k_3=1$ is the original background after a $\pi/2$-rotation.


\subsubsection*{Summary}

The various cases of T-duality transformations discussed in this section are summarized in 
table~\ref{tab_s3_t}. From there one can see how geometric charges and the NS-charge are interchanged,
and we highlight the cases of a globally-defined $U(1)_{\chi}$ and $U(1)_{\xi}$ isometry 
as well as the case $h=1$. We observe that, after a T-duality transformation along a globally-defined isometry of the three-sphere orbifold, the dual background is again a three-sphere orbifold.

\begin{table}[t]
\centering
\resizebox{\textwidth}{!}{
\renewcommand{\arraystretch}{2}
\begin{tabular}{l||ccc||ccc||ccc||ccc}
\multirow{2}{*}{model}&&&& \multicolumn{3}{|c||}{global $U(1)_{\chi}$} & \multicolumn{3}{|c||}{global $U(1)_{\xi}$}
& \multicolumn{3}{|c}{$h=1$}
\\[-6pt]
& $n_{\chi}$ & $n_{\xi}$ & $h$ & $n_{\chi}$ & $n_{\xi}$ & $h$ 
& $n_{\chi}$ & $n_{\xi}$ & $h$
& $n_{\chi}$ & $n_{\xi}$ & $h$  
\\ \hline\hline
$S^3$ & $\displaystyle\hphantom{-} \frac{k_1}{k_2}$ & $\displaystyle \hphantom{-}2\op \frac{k_2}{k_1}$ & $\displaystyle \hphantom{-}k_3$
& $\displaystyle k_1$ &  & $\displaystyle k_3$
&  & $\displaystyle 2\op k_2$ & $\displaystyle k_3$
& $\displaystyle\hphantom{-} \frac{k_1}{k_2}$ & $\displaystyle \hphantom{-}2\op \frac{k_2}{k_1}$ & $\displaystyle \hphantom{-}1$
\\
$T_{\chi}(S^3)$ & $\displaystyle \hphantom{-}k_3$ & $\displaystyle \hphantom{-}2\op \frac{1}{k_3}$ & $\displaystyle \hphantom{-}\frac{k_1}{k_2}$
& $\displaystyle k_3$ &  & $\displaystyle k_1$ 
 & &&
& $\hphantom{-}1$ & $\hphantom{-}2$ & $\displaystyle\hphantom{-} \frac{k_1}{k_2}$
\\
$T_{\xi}(S^3)$ & $\displaystyle \hphantom{-}\frac{1}{k_3}$ & $ \hphantom{-}\displaystyle 2\op k_3$ & $\displaystyle \hphantom{-}\frac{k_2}{k_1}$ 
& &&
& & $ \displaystyle 2\op k_3$ & $\displaystyle k_2$
& $\hphantom{-}1$ & $\hphantom{-}2$ & $\displaystyle\hphantom{-} \frac{k_2}{k_1}$
\\
$T_{(\chi,\xi)}(S^3)$ & $\displaystyle -\frac{k_2}{k_1}$ & $\displaystyle -2 \op \frac{k_1}{k_2}$ & $\displaystyle -\frac{1}{k_3}$ 
 &&&
 & &&
& $\displaystyle -\frac{k_2}{k_1}$ & $\displaystyle -2 \op \frac{k_1}{k_2}$ & $\displaystyle -1$  
\end{tabular}
}
\vspace*{10pt}
\caption{Summary of geometric and NS-charges for the three-sphere orbifold and its T-dual configurations. 
In the first column, $T_{\star}(S^3)$ denotes the T-dual of $S^3$ orbifold along the direction(s) $\star$.
For a globally-defined $U(1)_{\chi}$ isometry $k_2=1$ is needed, 
and we have displayed only the integer charges.
Similarly,  for a globally-defined 
$U(1)_{\xi}$ isometry $k_1=1$ has to be required and we again only showed the integer charges.
\label{tab_s3_t}
}
\end{table}


\subsection{T-duality along non-globally-defined $U(1)$ fibers}
So far we have discussed duality transformations for three-sphere orbifolds along the directions of the vector fields \eqref{eq:u1-vectors}. In this section, we analyze the local fields obtained by T-duality transformations along an arbitrary direction of the Clifford torus. In general, these directions will not be globally-defined $U(1)$ fibers, and consequently the dual backgrounds may not be globally-defined or may become non-compact.

For the present  discussion, we will go back to the original Hopf coordinates to avoid any choice of frame that singles out a particular direction. In this frame, a general sphere-orbifold takes the form
\eq{
\label{eq:Hopf-orbifold}
\arraycolsep2pt
\begin{array}{lcl}
ds^2&=& \displaystyle R^2 \op \left(\op d\eta^2+\frac{1}{\alpha_1^2}\cos^2\eta\,  d\xi_1^2+\frac{1}{\alpha_2^2}\sin^2\eta \, d\xi_2^2 \op\right) \,, \\[12pt]
H&=& \displaystyle 2\op \alpha_3 \sin\eta\,\cos\eta\, d\eta\wedge d\xi_1\wedge d\xi_2\,,
\end{array}
}
with $R^2=|\alpha_1\alpha_2\alpha_3|$ and $|\alpha_i| \in\mathbb Z_+$.  As follows from our previous discussion, not all values of $(\alpha_1,\alpha_2)$ are possible in order to have a globally-defined background. Next, we T-dualize along the isometry $\textsf{v}=\beta_1\partial_{\xi_1}+\beta_2\partial_{\xi_2}$ for arbitrary $(\beta_1,\beta_2)$ using the methods described in \cite{Plauschinn:2013wta,Plauschinn:2014nha}. After a choice of local coordinates $(\psi_1,\psi_2)$ we obtain the dual configuration
\eq{
\arraycolsep2pt
\begin{array}{lcl}
ds^2&=& \displaystyle R^2 \op \left(\op d\eta^2+\frac{\cos^2\eta}{\Delta}\,  d\psi_1^2+\frac{\sin^2\eta}{\Delta} \, d\psi_2^2 \op\right),\\[16pt]
H&=& \displaystyle \frac{\alpha_1^2\op\alpha_2^2\op\alpha_3 \op\beta_1\op\beta_2\sin(2\eta)}{\Delta^2}\,d\eta\wedge d\psi_1\wedge d\psi_2\,,\\[16pt]
e^{2\Phi}&=&\displaystyle e^{2\Phi_0} \op\frac{\alpha_1\alpha_2}{\alpha_3}\,\frac{1}{\Delta}\,,
\end{array}
}
where we have defined
\eq{ 
\Delta = \bigl( \alpha_1\op \beta_2\sin\eta\bigr)^2+\bigl( \alpha_2 \beta_1\cos\eta\bigr)^2 \,.
}
Since this is a local procedure, it does not give  information about the global properties of the coordinates $(\psi_1,\psi_2)$, which could be even 
non-compact. 
We however observe that by choosing $(\beta_1,\beta_2)=(\alpha_1,\pm\alpha_2)$ the dual configuration is locally a three-sphere, and for $\beta_1,\beta_2\neq 0$ the space is non-singular.


\section{NS5-branes}\label{sec:NS5-section}
\label{sec_ns5}

In this section we  discuss T-duality transformations along angular isometries for the 
NS5-brane solution and its orbifolds. 
T-duality transformations along the Hopf-fiber for the NS5-brane have been
considered in \cite{Duff:1998us,Duff:1998cr}, where the ten-dimensional solution was first dimensionally-reduced,
then a T-duality for the nine-dimensional solution was performed, which was then oxidized back to 
ten dimensions. Here we apply the Buscher rules directly, 
consider T-duality along a general direction, and discuss also orbifolds of the NS5-brane.
In addition to determining the T-dual configurations, 
we analyze the amount of supersymmetry preserved. 


\subsection{Prerequisites}
\label{sec_ns_pre}

To start, let us briefly give some details for the NS5-brane and its orbifolds. In particular, 
we determine the geometric and NS-charges.


\subsubsection*{NS5-brane}

The NS5-brane solution has a six-dimensional world-volume, and the 
four-di\-men\-sio\-nal transversal space can be described by the following field-configuration
\eq{\label{eq:NS5-brane}
\arraycolsep2pt
\begin{array}{lcl}
ds^2&=&\displaystyle h(r)\,dr^2+\frac{h(r)\,r^2}{4}\left(d\theta^2+d\xi^2+4d\chi^2-4\cos\theta d\chi d\xi\right),\\[8pt]
H&=&\displaystyle \star_4dh(r)\,,\\[8pt]
\displaystyle e^{2\Phi}&=&\displaystyle e^{2\phi_0} h(r)\,,
\end{array}
}
where $\star_4$ denotes the Hodge-star operator in four Euclidean dimensions and 
the variables take values 
$r\in[0,\infty)$, $\theta\in [0,\pi]$ and $\chi,\xi\in[0,2\pi)$.
The value of the dilaton at infinity $\phi_0$ is constant,
and the harmonic function $h(r)$ is given by
\begin{equation}\label{eq:h-function}
h(r)=1+\frac{k}{r^2}\,,
\end{equation}
where $k\in\mathbb Z_+$ is interpreted as the number of coincident NS5-branes. Note that the solution \eqref{eq:NS5-brane} can be seen as a three-sphere fibered along the radial direction.
This solution is asymptotically flat at $r\to\infty$, however, unlike empty Euclidean space the angular directions do not shrink  at the origin but the volume of the three-sphere at $r=0$ remains finite.


\subsubsection*{NS5-orbifolds}

In analogy to the case of the three-sphere, it is possible to construct a generalization of \eqref{eq:NS5-brane} by considering orbifold projections along the $U(1)$ fibers. The general form of this configuration is given by
\eq{\label{eq:NS5-orbifold}
\arraycolsep2pt
\begin{array}{lcl}
ds^2&=&\displaystyle h(r)\,dr^2+\frac{h(r)\,r^2}{4}\left(d\theta^2+\frac{1}{k_2^2}d\xi^2+\frac{4}{k_1^2}d\chi^2-\frac{4}{k_1 k_2}\cos\theta \,d\chi d\xi\right),\\[8pt]
H&=&\displaystyle \frac{k_3}{2}\sin\theta\,d\theta\wedge d\xi\wedge d\chi\,,\\[12pt]
\displaystyle e^{2\Phi}&=&\displaystyle e^{2\phi_0} h(r)\,,
\end{array}
}
where $|k_i|\in\mathbb{Z}_+$. The constraints on the possible values of $k_1$ and $k_2$ imposed by demanding a globally well-defined background coincide with those imposed on the three-sphere 
discussed in the last section. The harmonic function is now given by
\begin{equation}\label{eq:NS5-orbifold-hfunction}
h(r)=1+\frac{|k_1k_2k_3|}{r^2} \,,
\end{equation}
and this configuration is in general not asymptotically $\mathbb{R}^4$ anymore, but 
nevertheless a solution of the string equations-of-motion. The geometry at the origin $r=0$ 
is the  sphere-orbifold \eqref{eq:general-orbifold} with radius $R=\sqrt{|k_1k_2k_3|}$. 


\subsubsection*{Geometric charges}

Since for the above configurations the three-sphere 
formed by the angular directions does not shrink to zero at any point, 
we can describe  \eqref{eq:NS5-orbifold} as a  principal $U(1)$-bundle in the same way as
it was done for the ordinary three-sphere. Similarly as before, we can assign gauge fields to the fibration structures either along the coordinate $\chi$ or along $\xi$. In particular, for these two choices we have
\begin{equation}
\mathcal{A}_\chi=\frac{g_{\chi i}}{g_{\chi\chi}}\,dx^i\,,
\hspace{60pt}
\mathcal{A}_\xi=\frac{g_{\xi i}}{g_{\xi\xi}}\,dx^i\,,
\end{equation}
and by integrating the corresponding field strengths over a two-sphere surrounding the brane
at fixed radius
we determine the geometric charges as
\begin{equation}
n_{\chi}=\frac{k_1}{k_2}\,,
\hspace{60pt}
n_\xi=2\,\frac{ k_2}{k_1}\,.
\end{equation}
Note that $n_\chi\in\mathbb{Z}$ only when $k_2=1$ and $n_\xi\in 2\op\mathbb{Z}$ only when $k_1=1$, which 
are, respectively, the cases when the $U(1)_\chi$ and $U(1)_\xi$ fibers are globally-defined.


\subsubsection*{NS charges}

The NS charge $h$ of the above configuration can be determined by
integrating its Bianchi identity $dH = 4\pi^2 \op h\, \delta^{(4)}(x)$ 
over the transversal space
(see for instance \cite{Evslin:2008zm}),
which can be expressed as in integral of the $H$-flux over a three-sphere $S^3_{\infty}$  at $r=\infty$. 
The charge for the configuration \eqref{eq:NS5-orbifold} satisfies the quantization condition $h\in\mathbb{Z}$ and is given by
\eq{
  \label{ns_charge}
  h = \frac{1}{4\pi^2} \int_{\mathbb R^4} dH =\frac{1}{4\pi^2} \int_{S^3_{\infty}} H = k_3 \,.
}


\subsection{T-duality}
\label{sec_ns_dual}

We are now going to perform T-duality transformations for the NS5-brane background and its orbifolds. 
Similarly to the case of the three-sphere, after dividing by a $\mathbb{Z}_p$ action some of the isometries of the original NS5-brane might be broken. Applying T-duality along these directions leads to globally ill-defined configurations.


\subsubsection*{T-duality along the direction $\chi$}

We begin by studying T-duality along the direction $\chi$ in  \eqref{eq:NS5-orbifold}.
Applying the Buscher rules we find the following T-dual configuration
\eq{\label{eq:NS5dual-chi}
\arraycolsep2pt
\hspace{-20pt}\begin{array}{lcl}
ds^2&=&\displaystyle h(r)dr^2+\frac{h(r)\,r^2}{4} \left( d\theta^2+\frac{1}{k_2^2} \sin^2\theta d\xi^2 +\frac{k_1^2 }{h(r)^2\,r^4}(2 d\chi -k_3 \cos\theta \,d\xi)^2\right),\\[12pt]
B&=&\displaystyle -\frac{k_1}{2k_2}\cos\theta\, d\xi\wedge d\chi\,,    \\[14pt]
\displaystyle e^{2\Phi}&=&\displaystyle e^{2\phi_0}\,  r^{-2} \,, 
\end{array}\hspace{-20pt}
}
which is again a solution of the supergravity equations of motion. It is not asymptotically-flat anymore, and close to the origin the geometry is  locally a three-sphere.  With the same prescriptions used above we can assign the following charges to this background 
\begin{equation}
n'_\chi=k_3 = h\,,\hspace{50pt}h'=\frac{k_1}{k_2} = n_{\chi}\,.
\end{equation}
We observe that only in the case where $k_2=1$ the quantization condition $h'\in\mathbb{Z}$ will be satisfied. In fact, as it follows from the discussion on the three-sphere, this is the case where the $U(1)_\chi$ isometry of the original background is globally-defined. In this situation, the effect of T-duality along $\chi$ interchanges the geometric charge $n_\chi$ with the NS-charge $h$.


\subsubsection*{T-duality along the direction $\xi$}

As can be seen from \eqref{eq:NS5-orbifold}, we can equally-well perform a T-duality 
transformation  along the direction $\xi$. In analogy to the case of the three-sphere, the dual background is given by
\eq{\label{eq:NS5dual-xi}
\arraycolsep2pt
\hspace{-20pt}\begin{array}{lcl}
ds^2&=&\displaystyle h(r)dr^2+\frac{h(r)\,r^2}{4} \left( d\theta^2+\frac{4}{k_1^2} \sin^2\theta d\chi^2 +\frac{k_2^2 }{h(r)^2\,r^4}(d\xi +2k_3 \cos\theta d\chi)^2\right),\\[12pt]
B&=& \displaystyle\frac{k_2}{2k_1}\cos\theta\, d\xi\wedge d\chi\,,    \\[14pt]
\displaystyle e^{2\Phi}&=&\displaystyle e^{2\phi_0} \, r^{-2}\,,
\end{array}\hspace{-30pt}
}
which is again a non-asymptotically-flat solution to the string equations-of-motion. As in the case obtained by T-duality along $\chi$, one can compute the following charges
\begin{equation}
n'_\xi=-2\op k_3= -2\op h\,,\hspace{50pt}h'=-\frac{k_2}{k_1}= - \frac{1}{2}\op n_{\xi}\,.
\end{equation}
The resulting NS charge is an integer only in the case when $k_1=1$, which is the situation where the $U(1)_\xi$ fiber of the original background is globally-defined. We observe that, in this case, T-duality exchanges $k_3\leftrightarrow k_2$ as expected.


\subsubsection*{T-dualities along the directions $\chi$ and $\xi$}

Finally, we discuss the background obtained after two T-duality transformations along the directions
$\chi$ an $\xi$. 
(This is related to the $\beta$-deformation discussed in \cite{Lunin:2005jy}.)
The dual configuration can be expressed in the following way
\eq{
\label{eq:NS5-2Tdualities}
\arraycolsep2pt
\begin{array}{lcl}
ds^2&=& \displaystyle h(r)\left(dr^2+\frac{r^2}{4}\op d\theta^2\right)+
\frac{k_1^2\op k_2^2 \op r^2 \op h(r)}{4\op \Omega}\left(\frac{1}{k_1^2}\op d\xi^2+\frac{4 }{k_2^2}\op d\chi^2+
\frac{4}{k_1\op k_2}\cos\theta\op  d\xi\, d\chi\right),
\\[14pt]
B&=&\displaystyle \frac{k_1^2\op k_2^2\op k_3^2 \cos\theta}{2\op k_3 \op\Omega}\op d\xi\wedge d\chi\,,   
\\[14pt]
e^{2\Phi}&=&\displaystyle  e^{2\phi_0} \op \frac{k_1^2\op k_2^2 \op h(r)}{\Omega} \,,
\end{array}
\\[-12pt]
}
where we defined
\eq{
  \Omega = \bigl( r^2 h(r) \sin\theta\bigr)^2 + \bigl( k_1\op k_2\op k_3 \cos \theta \bigr)^2 \,.
}
As expected, this background is again a solution to the string equations-of-motion. 
However, the geometry is somewhat peculiar: at $r=0$ the directions $\theta$, $\xi$ and $\chi$ describe 
an $S^3$ orbifold, whereas at $r\to\infty$ the $\mathbb T^2$-fiber corresponding to $\xi$ and $\chi$ 
shrinks to zero size. The topology of the dual space is therefore different from the original NS5-brane topology. 
Furthermore, the NS-charge of the background \eqref{eq:NS5-2Tdualities} 
can in principle be computed in a similar ways as in \eqref{ns_charge}, but without 
proper knowledge of the dual topology this is difficult in practice. 
Finally, as we will discuss in the next section, \eqref{eq:NS5-2Tdualities} does not 
preserve any supersymmetry and hence the stability of this solution is not guaranteed.

On the other hand, we want to point-out that the configuration \eqref{eq:NS5-2Tdualities} obtained after 
two T-duality transformations does not show any non-geometric features -- contrary to 
what one might have naively expected from \cite{Shelton:2005cf}.
This is 
in contrast to compactifying the NS5-brane solution, smearing along the compact directions 
and performing a T-duality along the latter, after  which one obtains a non-geometric $5^2_2$-brane
\cite{deBoer:2012ma,Hassler:2013wsa}.


\subsection{Supersymmetry}
\label{sec_ns_susy}

We now want to analyze the amount of supersymmetry preserved by the
T-dual configurations determined in the last section. 
Even though the dual backgrounds are solutions to the string equations-of-motion, 
starting from the $1/2$-BPS NS5-brane solution we will see that a single T-duality along the direction 
$\chi$ or $\xi$ results in a $1/4$-BPS configuration. Furthermore, after two T-dualities supersymmetry 
is completely broken. These results are in agreement with \cite{Bakas:1994ba,Bergshoeff:1994cb,Bakas:1995hc,Alvarez:1995np}, where it was found that if a Killing spinor depends on 
the coordinate along which one T-dualizes then the corresponding supersymmetry will be broken.\footnote{A
formulation of this condition independent of a particular coordinate frame was given in 
\cite{Kelekci:2014ima} using the Kosmann spinorial Lie-derivative.}


\subsubsection*{Conventions}

In the rest of this section  we 
want to give some details of our analysis. 
For type II supergravity theories in ten dimensions 
the supersymmetry variations of the dilatini and gravitini read as follows 
(for our conventions see \cite{deBoer:2012ma})
\begin{equation}\label{eq:susy-variations}
\delta_\epsilon\lambda = \left(\frac{1}{2}\slashed\partial\Phi-\frac{1}{4}\slashed H \mathcal{P}\right)\epsilon, \hspace{30pt}\delta_\epsilon \Psi_M = \left(\nabla_M-\frac{1}{4}\slashed H_M\mathcal{P}\right)\epsilon ,
\end{equation}
where $\epsilon$ is a doublet of Majorana-Weyl spinors. The two components of $\epsilon$ have the same (IIB) or opposite (IIA) chiralities. The operator $\mathcal{P}$ acts on the spinor doublet as $\mathcal{P}=\sigma_3$ for type IIB and $\mathcal{P}=\Gamma_{[10]}\op\mathbb{I}_{2\times 2}$ for type IIA, where $\Gamma_{[10]}$ is the ten-dimensional chirality matrix. Note that if we choose a representation of eigenstates of $\Gamma_{[10]}$, $\mathcal{P}$ will act as $+\mathbb{I}_{32 \times 32}$ on one component of the doublet and as $-\mathbb{I}_{32 \times 32}$ on the other, both in type IIA/B. Therefore we will generically denote the two spinors as $\epsilon=(\epsilon_+,\epsilon_-)$.


\subsubsection*{The NS5-orbifold}
We begin analyzing the amount of supersymmetry preserved by the configuration \eqref{eq:NS5-orbifold}. This includes the NS5-brane for a particular choice of $k_1$ and $k_2$, which is known to be a $1/2$-BPS solution. The two Killing spinors for this background are\footnote{We note that $\epsilon_+$ is $2\pi k_1$-periodic in $\chi$ and 
$\epsilon_-$ is $4\pi k_2$-periodic in $\xi$.}
\eq{
\label{ks_ns5}
\arraycolsep2pt
\begin{array}{lclcl}
\epsilon_+&=&\displaystyle e^{\frac{\chi}{k_1}\Gamma_{\hat{7}}\Gamma_{\hat{8}}}\epsilon_{0,+}
&=& \displaystyle\left(\cos\frac{\chi}{k_1}+\Gamma_{\hat{7}}\Gamma_{\hat{8}}\sin\frac{\chi}{k_1}\right)\epsilon_{0,+}\,,
\\[14pt]
\epsilon_-&=&\displaystyle e^{-\frac{\theta}{2}\Gamma_{\hat{8}}\Gamma_{\hat{9}}}e^{\frac{\xi}{2k_2}\Gamma_{\hat{7}}\Gamma_{\hat{8}}}\epsilon_{0,-}
&=&\displaystyle \left(\cos\frac{\theta}{2}-\Gamma_{\hat{8}}\Gamma_{\hat{9}}\sin\frac{\theta}{2}\right)\left(\cos\frac{\xi}{2k_2}+\Gamma_{\hat{7}}\Gamma_{\hat{8}}\sin\frac{\xi}{2k_2}\right)\epsilon_{0,-}\,,
\end{array}
}
where $\epsilon_{0,\pm}$ are constant Majorana-Weyl spinors satisfying
$
(1\pm\Gamma_{\hat{6}}\Gamma_{\hat{7}}\Gamma_{\hat{8}}\Gamma_{\hat{9}})\epsilon_{0,\pm}=0
$,
which projects-out half of their components.\footnote{The solution presented here correspond to the case where $k_1k_2k_3>0$. Details corresponding to the opposite case can be found in appendix \ref{sec:susy-app}}


\subsubsection*{T-dual configurations}

Let us discuss the amount of supersymmetry preserved by the various T-dual configurations
discussed above. Here we only give the final results, but details of our computations
can be found in appendix \ref{sec:susy-app}.
\begin{itemize}

\item We start by considering the background \eqref{eq:NS5dual-chi} obtained after 
a T-duality transformation along the direction $\chi$.
Since $\epsilon_+$ in \eqref{ks_ns5} depends explicitly on $\chi$, we expect that 
the corresponding supersymmetry will be broken. 
Indeed, for \eqref{eq:NS5dual-chi} we find only one Killing spinor given by (assuming $k_1k_2k_3>0$)
\eq{
\epsilon_- &=e^{-\frac{\theta}{2}\Gamma_{\hat{8}}\Gamma_{\hat{9}}}e^{\frac{\xi}{2k_2}\Gamma_{\hat{7}}\Gamma_{\hat{8}}}\epsilon_{0,-}\\
& =\left(\cos\frac{\theta}{2}-\Gamma_{\hat{8}}\Gamma_{\hat{9}}\sin\frac{\theta}{2}\right)\left(\cos\frac{\xi}{2k_2}+\Gamma_{\hat{7}}\Gamma_{\hat{8}}\sin\frac{\xi}{2k_2}\right)\epsilon_{0,-}\,,
}
where again $(1-\Gamma_{\hat{6}}\Gamma_{\hat{7}}\Gamma_{\hat{8}}\Gamma_{\hat{9}})\epsilon_{0,-}=0$.
Note that this configuration preserves only half of the original supersymmetries.

\item A similar analysis applies to a T-duality transformation along the direction $\xi$. 
Since in \eqref{ks_ns5} the Killing spinor $\epsilon_-$ depends explicitly on $\xi$, 
we expect that the corresponding supersymmetry will be broken under T-duality. 
Indeed, for the background \eqref{eq:NS5dual-xi} we find only one Killing spinor given by  (assuming $k_1k_2k_3>0$)
\eq{
\epsilon_+&=e^{-\frac{\theta}{2}\Gamma_{\hat{8}}\Gamma_{\hat{9}}}e^{\frac{\chi}{k_1}\Gamma_{\hat{7}}\Gamma_{\hat{9}}}\epsilon_{0,-}
\\
&=\left(\cos\frac{\theta}{2}-\Gamma_{\hat{8}}\Gamma_{\hat{9}}\sin\frac{\theta}{2}\right)\left(\cos\frac{\chi}{k_1}+\Gamma_{\hat{7}}\Gamma_{\hat{9}}\sin\frac{\chi}{k_1}\right)\epsilon_{0,+}\,,
}
with $
(1+\Gamma_{\hat{6}}\Gamma_{\hat{7}}\Gamma_{\hat{8}}\Gamma_{\hat{9}})\epsilon_{0,+}=0 
$. This configuration again again preserves only half of the original supersymmetries.

\item After applying two T-duality transformations along the directions $\chi$ and $\xi$, 
we expect that both of the supersymmetries corresponding to $\epsilon_+$ and 
$\epsilon_-$ in \eqref{ks_ns5} will be broken. 
For the background  \eqref{eq:NS5-2Tdualities} the supersymmetry variations 
\eqref{eq:susy-variations} can only  be solved for vanishing spinors, and hence
supersymmetry is completely broken. 
This means that stability is no longer guaranteed, and therefore we 
do not study this background in more detail.

\end{itemize}


\subsection{T-duality along non-globally-defined $U(1)$ fibers}

We finally want to generalize our previous discussion in the following way: 
if we interpret the three-sphere inside the transversal geometry of the NS5-brane solution as
a two-torus fibered over a line-segment, we can in principle perform 
T-duality transformations also along an arbitrary direction of the two-torus. 
In general, such isometries are not globally well-defined and hence the dual 
background may show global problems. Nevertheless, locally this analysis is valid.

Let us rewrite the NS5-brane solution \eqref{eq:NS5-brane} in a different set of coordinates
which make the $\mathbb T^2$-fibration structure explicit. Including 
orbifold projections along the two directions of the two-torus, we have 
\eq{
\label{eq:Hopf-NS5orbifold}
\arraycolsep2pt
\begin{array}{lcl}
ds^2&=& \displaystyle h(r) \op \left(\op dr^2 +r^2\op d\eta^2+\frac{r^2}{\alpha_1^2}\cos^2\eta\,  d\xi_1^2+\frac{r^2}{\alpha_2^2}\sin^2\eta \, d\xi_2^2 \op\right) \,, \\[8pt]
H&=& \displaystyle 2\op \alpha_3 \sin\eta\,\cos\eta\, d\eta\wedge d\xi_1\wedge d\xi_2\,,\\[8pt]
e^{2\Phi}&=&e^{2\phi_0}\op h(r) \,,
\end{array}
}
where $\eta\in[0,\pi/2]$ and $\xi_{1,2}\in[0,2\pi)$ and $|\alpha_i|\in\mathbb Z_+$, and where the harmonic function is given by
\begin{equation}
h(r)=1+\frac{|\alpha_1\alpha_2\alpha_3|}{r^2} \,.
\end{equation}
After performing a T-duality transformation along the direction $\textsf{v}=\beta_1\partial_{\xi_1}+\beta_2\partial_{\xi_2}$ we obtain, for a choice of local coordinates $(\psi_1,\psi_2)$, the configuration
\eq{
\label{dual_gen_001}
\resizebox{\textwidth}{!}{$
\arraycolsep2pt
\begin{array}{lcl}
ds^2&=& \displaystyle h(r)\left(dr^2+r^2d\eta^2+\frac{r^2}{4}\,\frac{\sin^2(2\eta)}{\Delta}\,d\psi_1^2\right)
+\frac{1}{r^2h(r)}\frac{\alpha_1^2\alpha_2^2}{\Delta}\left(d\psi_2-\frac{\alpha_3}{2}\cos(2\eta)\op d\psi_1\right)^2,
\\[14pt]
H&=&\displaystyle \frac{\beta_1\beta_2\alpha_1^2\alpha_2^2\,\sin^2(2\eta)}{\Delta^2}\,d\eta\wedge d\psi_1\wedge d\psi_2\,,
\\[14pt]
e^{2\Phi}&=& \displaystyle e^{2\phi_0}\op \frac{\alpha_1^2\op\alpha_2^2}{r^2\Delta}\,,
\end{array}
$}\\[-8pt]
}
where we defined
\eq{ 
\Delta = \bigl( \alpha_2\op \beta_1\sin\eta\bigr)^2+\bigl( \alpha_1 \beta_2\cos\eta\bigr)^2 \,.
}
As we mentioned before, following Buscher's procedure in general does not give 
global information about the T-dual space, and hence we have not specified the range  of the coordinates $(\psi_1,\psi_2)$. Locally, however, \eqref{dual_gen_001} does solve the string equations-of-motion. 
Concerning the amount of supersymmetry preserved by \eqref{dual_gen_001}, 
for arbitrary $(\beta_1,\beta_2)$ all supersymmetries are broken -- only for $(\beta_1,\beta_2)=(\alpha_1,\pm\alpha_2)$ the solution preserves half of the original supersymmetries.
Note that the latter are precisely the examples \eqref{eq:NS5dual-chi} and \eqref{eq:NS5dual-xi}
discussed above. 


\section{Discussion}
\label{disc}

In this note we have studied T-duality transformations along angular directions for the 
NS5-brane  and its orbifolds. 
This solution has a six-dimensional world-volume, and the  four-dimensional space transversal to the brane can be seen as a three-sphere fibration
along a radial direction. 
\begin{itemize}

\item In section~\ref{sec_three_sphere} we therefore first analyzed T-duality transformations for the three-sphere
along one and two directions. We paid special attention to global issues, 
discussed orbifolds of the three-sphere, determined corresponding geometric charges and 
computed the $\beta$-transform of the three-sphere.

\item In section~\ref{sec_ns5} we then applied our findings to T-duality transformations of the NS5-brane
and its orbifolds. We computed the NS- and geometric charges, we determined the background after one and two T-dualities along 
angular directions, and we analyzed the amount of supersymmetry for the dual solutions. 
In agreement with expectations from the  literature  \cite{Bakas:1994ba,Bergshoeff:1994cb,Bakas:1995hc,Alvarez:1995np}, we found that one T-duality preserves at most one-half of the original supersymmetry while two 
T-dualities break supersymmetry completely. 
We furthermore observed that after two T-dualities the background is geometric, 
contrary to what one might have expected from \cite{Shelton:2005cf}.

\end{itemize}
We now want to discuss in some more detail the implications of our results.


\subsection{Comparison with T-duality for $\mathbb R^n$}

Let us note that performing a T-duality transformation along an angular direction 
for empty Euclidean space $\mathbb R^n$ results in a dual geometry which is singular at the origin \cite{Rocek:1991ps,Adams:2001sv}. The reason is that the norm of the corresponding Killing vector vanishes there. 
This is a puzzling observation, and it has been suggested that winding modes and 
world-sheet instantons may play
a role in resolving the singularity.
On the other hand, we see for instance from \eqref{eq:NS5-orbifold} that the metric and $H$-flux 
of the NS5-brane at the origin $r=0$ is finite. (We ignore the dilaton in the present discussion.)
Performing a single T-duality transformation 
along an angular direction leads to a geometry which is again non-singular, as can be seen 
for instance from \eqref{eq:NS5dual-chi}, and
the reason for the non-singular behavior of the dual metric at $r=0$ can be traced back to the non-vanishing
$H$-flux $h=k_3\neq0$.

We do not have an answer to the question whether or how the singularity of the T-dual of $\mathbb R^n$ 
can be resolved. However, for the NS5-brane the $H$-flux plays an important role in regard to 
this point -- and therefore one might suspect that also for the 
case of $\mathbb R^n$
a non-trivial $B$-field or $H$-flux has to be included. 
We are planning to come back to this point in the future.


\subsection{Comparison with toroidal compactifications}

We also want to compare our results to T-duality transformations for the compactified 
NS5-brane solution. As we mentioned in the introduction, in this approach 
one breaks the isometry of the  NS5-brane in the transversal space 
from $\mathfrak{so}(4)$ to $\mathfrak{so}(4-n)$
by compactifying 
on a $n$-torus $\mathbb T^n$ with $n=1,2,\ldots$
In a second step the localized solution is smeared along the compact directions which 
enhances the isometry algebra
\eq{
  \mathfrak{so}(4) \;\longrightarrow\;
  \mathfrak{so}(4-n) \;\longrightarrow\;
  \mathfrak{so}(4-n) \times \bigl[ \mathfrak{u}(1) \bigr]^n \,.
}
T-duality transformations for these configurations are performed along the compact directions 
with $\mathfrak{u}(1)$ isometries, leading for instance to the KK-monopole and the 
$5^2_2$-brane. Note that in these cases T-duality preserves supersymmetry.


\subsubsection*{Geometric flux}

Let us emphasize that when compactifying the NS5-brane, the non-compact trans\-ver\-sal space
 can no longer be described as a three-sphere fibration along a radial direction. Furthermore, 
the geometric charge characterizing the non-triviality of the fibration vanishes, 
and hence, in some sense, the compactification breaks the geometric charge. 
This is of course not a problem, but what our work shows is that 
the original NS5-solution and the compactified-and-smeared solution have different topological properties.


\subsubsection*{Comparison to DFT}

The chain of T-duality transformations between the smeared NS5-brane, KK-monopole 
and $5^2_2$-brane mentioned in the introduction has also been discussed 
in double field theory  (see for instance \cite{Berman:2014jsa,Bakhmatov:2016kfn,Blair:2017hhy}). 
It would be interesting to also incorporate T-duality transformations along 
angular directions into this framework  -- which should be possible if DFT is a background-independent
formulation. 
On the other hand, as we have verified for the NS5-brane, T-duality can break supersymmetries
\cite{Bakas:1994ba,Bergshoeff:1994cb,Bakas:1995hc,Alvarez:1995np}
and therefore the generators of supersymmetry and T-duality transformations
do not need to commute. It would be interesting to see if a supersymmetric formulation of 
DFT can be made compatible with this observation.


\subsubsection*{The NS5--Taub-NUT configuration}

We also want to give some more technical details on the difference between the 
NS5-backgrounds studied in this work and the compactified-and-smeared configurations.

To obtain the latter, one starts from the NS5-brane solution \eqref{eq:NS5-brane},
splits the transversal space as $\mathbb R^4\to\mathbb R^3\times \mathbb R$,
introduces $\rho$ as the radial coordinate in $\mathbb R^3$ and $x$ as the coordinate in $\mathbb R$,
and compactifies the the $x$-direction on a circle. 
For the harmonic function one then finds \cite{Gregory:1997te}
\begin{equation}
h(r)=1+\sum_{n\in\mathbb{Z}}\frac{k}{\rho^2+(x+2\pi n)^2}
\hspace{30pt}
\xrightarrow{\quad\rho\gg 1\quad}
\hspace{30pt}
1+\frac{k}{2\rho}=\mathsf{h}(\rho)\,,
\end{equation}
where $\rho\gg1$ corresponds to the smearing limit. 
In this limit, the background no longer depends on $x$ and a corresponding direction of isometry appears.
Note, however, that by following this procedure the topology of the background has been changed. More concretely, the $U(1)$ along the coordinate $x$ is now trivially-fibered and has vanishing geometric charge, so the 
NS5-background characterized by the smeared function $\mathsf{h}(r)$ has only an NS-charge $h=k$. 
The smearing approximation breaks down at the limit $\rho\rightarrow 0$, and one needs to include corrections that reproduce the original function $h(r)$. These corrections re-introduce a dependence on the compact coordinate $x$ which breaks the isometry, and such corrections can be understood as instanton corrections \cite{Tong:2002rq}.

By performing a T-duality transformation along the compact direction $x$, one obtains a Taub-NUT space (also called a KK-monopole) with no NS-charge and $k$ units of geometric charge \cite{Gross:1983hb,Sorkin:1983ns,Bossard:2008sw}. 
With this knowledge, we can construct geometries with both non-trivial NS- and geometric charges by putting $k_1$ KK-monopoles and $k_3$ smeared NS5-branes together. The resulting configuration is
\eq{
\label{smear_ns_kk}
\arraycolsep2pt
\begin{array}{lcl}
ds^2&=&\displaystyle \mathsf{h}_1(\rho) \mathsf{h}_3(\rho)\bigl(d\rho^2+\rho^2d\theta^2+\rho^2\sin^2\theta d\varphi^2\bigr)+\frac{\mathsf{h}_3(\rho)}{\mathsf{h}_1(\rho)}\left(dx-\frac{k_1}{2} \cos\theta d\varphi\right)^2\,,\\[8pt]
B&=&\displaystyle  -\frac{k_3}{2}\cos\theta\, d\varphi\wedge dx\,,    \\[12pt]
\displaystyle  e^{2\Phi}&=&\displaystyle e^{2\phi_0}\op \mathsf{h}_3(\rho)\,,
\end{array} \hspace*{-5pt}
}
with $\rho\in[0,\infty)$, $\theta\in[0,\pi]$ and $\varphi,x\in[0,2\pi)$. The harmonic functions $\mathsf{h}_1(\rho)$ and $\mathsf{h}_3(\rho)$ correspond to the KK-monopole and the smeared NS5-brane, respectively, and read
\begin{equation}
\mathsf{h}_i(\rho)=1+\frac{k_i}{2\op \rho} \,.
\end{equation}
A computation analogous to the ones carried before shows that this background has $k_1$ units of geometric charge and $k_3$ units of NS-charge. 
After applying T-duality to the direction $x$, the resulting configuration is again a superposition of NS5-branes and KK-monopoles, where the numbers $k_1$ and $k_3$ have been interchanged
\eq{
  k_1 \quad \xleftrightarrow{\hspace{10pt}\mbox{\scriptsize T-duality}\hspace{10pt}}
  \quad k_3 \,.
}  
We furthermore note that in the limit $\rho\rightarrow 0$ the metric in \eqref{smear_ns_kk} remains finite, 
and the corresponding geometry is given by the three-sphere orbifold $SU(2)_{k_1k_3}/\mathbb{Z}_{k_1}$ \cite{Maldacena:2001ky}. 
However, even though the solution \eqref{smear_ns_kk} close to the origin agrees with 
the $r\to0$ behavior of the ones discussed in section~\ref{sec_ns5}, away from $r=0$ they are different.


\vspace{0.5cm}
\subsubsection*{Acknowledgments} 

We thank Ismail Achmed-Zade, Stefano Massai, Christoph Mayrhofer and Alexander Tabler for helpful discussions. 
We furthermore thank the participants of the 
workshop ``String dualities and geometry'' in Bariloche for useful comments, where 
part of this work has been presented.


\clearpage
\appendix

\section{Hopf fibration and $U(1)$ actions}\label{sec:Hopf}
\label{app_hopf}

The three-sphere $S^3$ can be described as a non-trivial $S^1$ fibration  
as 
\begin{equation}
S^1\lhook\joinrel\longrightarrow S^3\overset{\pi}{\longrightarrow} S^2 \,,
\end{equation}
and the standard way is by using the projection
\eq{
\pi : \hspace{5pt}
\left\{
\begin{array}{ccl}
S^3&\longrightarrow & \hspace{5pt}S^2 \\[4pt]
(z_0,z_1)&\longmapsto & (2 z_0z_1^*,\,|z_0|^2-|z_1|^2) 
\end{array}\right.
}
It is easy to check that $\pi(z_0,z_1)\in S^2$ and that $\pi(z_0,z_1)=\pi(w_0,w_1)$ if an only if $(z_0,z_1)=(\lambda w_0, \lambda w_1)$ for $\lambda\in U(1)$. Therefore, at each point of $S^2$ 
there is  a $U(1)$ fiber. 
Note that the action of $U(1)$ on each fiber is transitive 
and free. 
The fibration is then a principal $U(1)$ bundle.
To define the standard Hopf fibration, we use the first expression in \eqref{eq:u1 actions} to construct the $U(1)$ action on the fiber. 
However, we can also employ the second action in \eqref{eq:u1 actions}  as
\eq{
\bar\pi : \hspace{5pt}
\left\{
\begin{array}{ccl}
S^3&\longrightarrow & \hspace{5pt}S^2 \\[4pt]
(z_0,z_1)&\longmapsto & (2 z_0z_1,\,|z_0|^2-|z_1|^2) 
\end{array}\right.
}
Again, one can show that  $\bar\pi(z_0,z_1)\in S^2$ and that, now, $\bar\pi(z_0,z_1)=\bar\pi(w_0,w_1)$ if an only if $(z_0,z_1)=(\lambda w_0, \lambda^* w_1)$ for $\lambda\in U(1)$. As before, the action acts transitively and freely on all fibers, and the fibration is again a principal $U(1)$ bundle.

Let us also address the question whether one can construct fibrations using  actions infinitesimally characterized by a linear combination of the vectors $v$ and $\bar{v}$ in \eqref{eq:u1-vectors}. If we take the linear combination $a\op v+ b \op\bar v$ with  where $a,b\in\mathbb R$, the corresponding group action is
\begin{equation}\label{eq:general-u1-action}
g(z_0,z_1)=(e^{i (a+b) \alpha} z_0,\,e^{i (a-b)\alpha}z_1) \,,
\end{equation}
where $\alpha\in [0,2\pi)$. 
Consider now a $U(1)$ orbit through the point $(z_0,z_1)=(0,1)$ defined as all points satisfying $|z_1|=1$ and $z_0=0$. The action \eqref{eq:general-u1-action} acts freely and transitively on it if and only if $a+b=\pm 1$. In other words, only when this condition is satisfied the orbit through the point $(z_0,z_1)=(0,1)$ constructed with the action \eqref{eq:general-u1-action} is isomorphic to $U(1)$.  Similarly, the action is free and transitive on the $U(1)$ orbit through the point $(z_0,z_1)=(0,1)$ if and only if $a-b=\pm 1$. Then, we can conclude that the only cases that can be used to construct principal bundles are $(a,b)=(\pm 1,0),\,(0,\pm 1)$, which are the ones 
discussed above.

Finally, we note  that although the three-sphere $S^3$ can be described as a principal $U(1)$ bundle in two different ways, it cannot be described as a principal $U(1)\times U(1)$ bundle. The reason is that the orbits constructed using the $U(1)\times U(1)$ action
\begin{equation}
g(z_0,z_1)=(e^{i (\alpha+\beta)} z_0,\,e^{i (\alpha-\beta)}z_1) \,,
\end{equation}
with $\alpha,\beta\in [0,2\pi)$, are not isomorphic to $U(1)\times U(1)$ everywhere. In particular, the orbits through the points $(z_0,z_1)=(0,1)$ and $(z_0,z_1)=(1,0)$ are two $U(1)$ one on top of each other.


\section{Lens spaces}\label{sec:Lens}

Given two natural numbers $p$ and $q$ with $p>q$ and relatively prime, the lens space $L(p,q)$ is defined as the orbifold $S^3/\mathbb{Z}_{p}$ constructed with the action
\begin{equation}\label{eq:lens-space-action}
(z_0,z_1)\rightarrow\left(e^{\frac{2\pi i m}{p}}z_0,\,e^{\frac{2\pi i\, q\, m }{p}}z_1\right),
\end{equation}
for any $m\in \mathbb{Z}_p$. This action is free if and only if the condition of $p$ and $q$ being relatively prime is satisfied. Note that, in fact, $q$ is defined modulo $p$. In terms of Hopf coordinates \eqref{eq:hopf-coordinates}, the action is
\begin{equation}\label{eq:orbifold-action}
\xi_1\rightarrow\xi_1+\frac{2\pi m}{p}\,,
\hspace{60pt}
\xi_2\rightarrow\xi_2+\frac{2\pi q\op m}{p}\,.
\end{equation}


\subsubsection*{The $\mathbb{Z}_{k_1}^{(\chi)}$ orbifold}

Taking $q=1$, the action \eqref{eq:orbifold-action} leave the coordinate $\xi$ invariant and acts on $\chi$ as
\begin{equation}
\chi\rightarrow\chi+\frac{2\pi m}{p}\,.
\end{equation}
Then, the configuration \eqref{eq:orbifold-chi} where we constructed the orbifold by considering the $\mathbb{Z}_{k_1}$ action on $\chi$ is the space $L(k_1,1)$.


\subsubsection*{The $\mathbb{Z}_{k_2}^{(\xi)}$ orbifold}

The action that leaves $\chi$ invariant and acts only on $\xi$ corresponds to the case $q=p-1$ (equivalent to $q=-1$). However, in this case the situation is more subtle. 
The reason is that, as discussed before \eqref{eq:geometric-charge-xi}, the coordinates \eqref{eq:modified-Hopf-coordinates} are not the most appropriate to describe the three-sphere as a $U(1)_\xi$ fibration. In particular, if one considers the inverse transformation of \eqref{eq:modified-Hopf-coordinates},
\begin{equation}
\xi_1=\chi+\frac{1}{2}\op\xi\,,\hspace{60pt}\xi_2=\chi-\frac{1}{2}\op\xi \,,
\end{equation} 
one can easily see that, by sending $\xi\rightarrow\xi+2\pi$, the coordinates $\xi_1$ and $\xi_2$ do not come back
to themselves. The intuition behind this fact is that, when assigning a period of $2\pi$ to $\xi_1-\xi_2$, we are effectively orbifolding this direction. In fact, the points related by the $\mathbb{Z}_2$ action
\begin{equation}
(z_0,z_1)\rightarrow\left(e^{i \pi m}z_0,e^{- i\pi m}z_1\right),
\end{equation}
with $m\in\{0,1\}$, are identified in the coordinates $(\xi,\chi)$. We stress, however, that this is an artifact of the coordinate frame we use, and not an orbifolding of the actual background. Nevertheless, this effect arises when computing the charges, since we compute them in a particular frame. This is then the reason for the factor $4\pi$  in the normalization of \eqref{eq:geometric-charge-xi} and the even charge along this direction. Changing to the coordinate frame \eqref{eq:modified-Hopf-coordinates-tilde} will move this effect from the direction $\xi$ to the direction $\tilde{\chi}$.

With this considerations, we next want to find an action that leaves $\chi$ invariant. The natural guess is
\begin{equation}\label{eq:orbifold-xi-action}
(z_0,z_1)\rightarrow\left(e^\frac{2\pi i m_2}{ k_2}z_0,e^{-\frac{2\pi i m_2}{k_2}}z_1\right),
\end{equation}
for $m_2\in\mathbb{Z}_{k_2}$. The space constructed using this action is the $L(k_2,k_2-1)$ lens space. In terms of the coordinate system \eqref{eq:modified-Hopf-coordinates}, the action \eqref{eq:orbifold-xi-action} acts on the coordinate $\xi$ as
\begin{equation}\label{eq:orbifold-action-xi}
\xi\rightarrow\xi+\frac{4\pi m_2}{k_2} \,.
\end{equation}
Note that, for even $k_2$, the action looks rather as a $\mathbb{Z}_{k_2/2}$-orbifold. However, since $m_2\in\mathbb{Z}_{k_2}$, the action \eqref{eq:orbifold-action-xi} has then fixed points and the corresponding configuration fails to be an appropriate description for the orbifold constructed with the action \eqref{eq:orbifold-xi-action}.  Again this is just a coordinate dependent effect. In fact, lens spaces $L(k_2,k_2-1)$ with even $k_2$ can be described using the coordinates \eqref{eq:modified-Hopf-coordinates-tilde}. In this frame, $L(k_1,1)$ spaces can only be described in the cases where $k_1$ is odd.


\subsubsection*{The $\mathbb{Z}_{k_1}^{(\chi)}\times\mathbb{Z}_{k_2}^{(\xi)}$ orbifold}
Finally, under certain assumptions it is possible to construct orbifolds with an action along both  $\chi$ and $\xi$ directions. The natural generalization to the orbifold actions discussed above is
\begin{equation}\label{eq:z2-orbifold-action}
(z_0,z_1)\rightarrow\left(e^{2\pi i\frac{k_2 m_1+k_1 m_2}{k_1  k_2}}z_0,e^{2\pi i\frac{ k_2 m_1-k_1 m_2}{k_1  k_2}}z_1\right),
\end{equation}
with $m_i\in\mathbb{Z}_{k_i}$. In order to have a globally well-defined orbifold, one has to check whether this action is free. As already discussed, the two $U(1)$ isometries of the three-sphere collapse into a single $U(1)$ in two points of the base. However, if the integers $k_1$ and $k_2$ are relatively prime, the action \eqref{eq:z2-orbifold-action} becomes free at these points. In particular, one can convince oneself that the sets $\{\tilde k_2 m_1+k_1 m_2\,|\,m_1\in [0,k_1-1],m_2\in [0,\tilde k_2-1]\}$ and $\{\tilde k_2 m_1-k_1 m_2\,|\,m_1\in [0,k_1-1],m_2\in [0,\tilde k_2-1]\}$ 
 contain exactly the same elements as $\mathbb{Z}_{k_1 k_2}$ but in a different order. Therefore, the total space will be $L(k_1 k_2,q)$ for some $q\neq 1$, and the resulting space is in fact a $\mathbb{Z}_p$-orbifold  along an oblique direction.  

In terms of the coordinates \eqref{eq:modified-Hopf-coordinates}, the action \eqref{eq:z2-orbifold-action} is
\begin{equation}
\chi\rightarrow\chi+\frac{2 \pi m_1}{k_1}\,,
\hspace{60pt} 
\xi\rightarrow\xi+\frac{4 \pi m_2}{ k_2} \,,
\end{equation} 
where again the frame fails to describe the cases with even $k_2$. These cases can be described using the coordinates \eqref{eq:modified-Hopf-coordinates-tilde}, which fail to describe cases with even $k_1$. Note that $k_1$ and $k_2$ can never be even at the same time.


\section{Supersymmetry analysis}\label{sec:susy-app}

In this appendix we give details of the analysis of the supersymmetry variations \eqref{eq:susy-variations} for some of the configurations appearing in section \ref{sec:NS5-section}. During the calculation the following notation will be used: 
\begin{itemize}
\item Capital letters $M,N,...$  correspond to curved space-time indices and take values  $M\in\{0,\dots 9\}$.
\item Space-time indices are separated into those along the brane, $\mu\in\{0,\dots,5\}$, and those perpendicular to it, $i\in\{r,\theta,\chi,\xi\}$. 
\item Flat indices are denoted by a hat. For the transverse space we have $\hat{\imath}\in\{\hat{6},\hat{7},\hat{8},\hat{9}\}$.
\end{itemize}
The ten-dimensional fields are constructed by trivially adding the brane-volume directions to the fields describing the transverse space, like \eqref{eq:NS5-brane}. Note that both the spin-connection and the NS field strength will have non-zero components only along the directions transversal to the brane.
Furthermore, we use conventions where $\slashed H=\frac{1}{3!}\Gamma^{\hat{M}\hat{N}\hat{P}}H_{\hat{M}\hat{N}\hat{P}}$, $\slashed H_M=\frac{1}{2!}\Gamma^{\hat{N}\hat{P}}H_{\hat{N}\hat{P}M}$ and $\Gamma^{\hat{M}\dots\hat{N}}=\frac{1}{p!}\Gamma^{[\hat{M}}\dots\Gamma^{\hat{N}]}$.


\subsubsection*{The NS5-orbifold}

We will first analyze the supersymmetry preserved by the NS5-orbifold solution \eqref{eq:NS5-orbifold}. This includes the NS5-brane as a particular case, which is well-known that it preserves half of the supersymmetries.


\paragraph{Dilatino variation}
We begin by analyzing the first of the variations in \eqref{eq:susy-variations}. A direct calculation shows that, in the present case,
\begin{equation}
\delta_\epsilon\lambda=-\frac{|k_1 k_2 k_3|}{2 r^3 h(r)^{3/2}}\,\Gamma_{\hat{6}}\, (1\pm\text{sgn}(k_1 k_2 k_3)\Gamma _{\hat{6}}\Gamma _{\hat{7}} \Gamma _{\hat{8}} \Gamma _{\hat{9}})\,\epsilon \,,
\end{equation}
where $\text{sgn}(x)$ is the sign function and we pick the plus sign when acting on $\epsilon_+$ and the minus when acting on $\epsilon_-$. The matrix $\Gamma_{[4]}=\Gamma _{\hat{6}}\Gamma _{\hat{7}} \Gamma _{\hat{8}} \Gamma _{\hat{9}}$ is the chirality operator of a representation of the four dimensional euclidean Clifford algebra and, therefore, $\frac{1}{2}\left(1\pm\Gamma_{[4]}\right)$ are projectors. Also, note that the two chirality operators $\Gamma_{[4]}$ and $\Gamma_{[10]}$ commute and one can construct spinors which are chiral with respect to both of them. We conclude then, that supersymmetry variations of the dilatino vanishes for spinors satisfying the condition. 
\begin{equation}\label{eq:projection-condition-orbifold}
\bigl(1\pm\text{sgn}(k_1 k_2 k_3)\Gamma _{\hat{6}}\Gamma _{\hat{7}} \Gamma _{\hat{8}} \Gamma _{\hat{9}}\bigr)\,\epsilon_\pm=0\,,
\end{equation}
which reduces the degrees of freedom of the original Majorana-Weyl spinors by one half. For the following we will assume that without loss of generality $\text{sgn}(k_1 k_2 k_3)=1$. In the case where $\text{sgn}(k_1 k_2 k_3)=-1$, the two spinors of the doublet $\epsilon$ are interchanged.


\paragraph{Gravitino variation for $\epsilon_+$}
We next analyze the second condition in \eqref{eq:susy-variations} for the component $\epsilon_+$. For the components $M=\mu$, the equations $\delta_\epsilon\Psi_M=0$ reduce to $\partial_\mu \epsilon=0$, and the Killing spinors have to be constant along the brane directions. For $M=i$ the variations are 
\eq{
\arraycolsep2pt
\begin{array}{lcl}
\delta_{\epsilon_+}\Psi_r&=&\partial_r\epsilon_+,
\\[4pt]
\delta_{\epsilon_+}\Psi_\theta&=&\partial_\theta\epsilon_+ +\frac{1}{4 h(r)}\left(\Gamma _{\hat{8}} \Gamma_{\hat{9}}-\Gamma _{\hat{6}} \Gamma _{\hat{7}}\right)\epsilon_+ \,,
\\
\delta_{\epsilon_+}\Psi_\xi&=&\partial_\xi\epsilon_+-\frac{1}{4k_2 h(r)}\Big(\sin \theta \left(\Gamma _{\hat{7}} \Gamma _{\hat{9}}+\Gamma _{\hat{6}} \Gamma _{\hat{8}} \right)+\cos \theta \left(\Gamma _{\hat{7}}\Gamma _{\hat{8}} -\Gamma_{\hat{6}}\Gamma _{\hat{9}} \right)\Big)\epsilon_+\,,
\\
\delta_{\epsilon_+}\Psi_\chi&=&\partial_\chi\epsilon_+-\frac{1}{2 r^2k_1 h(r)}\Big(2 k_1k_2k_3\,\Gamma _{\hat{7}} \Gamma _{\hat{8}} +r^2\left(\Gamma _{\hat{7}} \Gamma _{\hat{8}}+\Gamma _{\hat{6}} \Gamma _{\hat{9}} \right)\Big)\epsilon_+\,, 
\end{array} 
}
and applying them to a spinor $\epsilon_+$ satisfying \eqref{eq:projection-condition-orbifold} they reduce to
\eq{
\arraycolsep2pt
\begin{array}{lcl@{\hspace{50pt}}lcl}
\delta_{\epsilon_+}\Psi_r&=&\partial_r\epsilon_+\,,
&
\delta_{\epsilon_+}\Psi_\xi&=&\partial_\xi\epsilon_+\,,
\\[4pt]
\delta_{\epsilon_+}\Psi_\theta&=&\partial_\theta\epsilon_+\,,
&
\delta_{\epsilon_+}\Psi_\chi&=&\bigl(\partial_\chi-\tfrac{1}{k_1}\Gamma _{\hat{7}} \Gamma _{\hat{8}}\bigr)\epsilon_+
\,.
\end{array} 
}
A general solution to the equations $\delta_{\epsilon_+}\lambda=0$ and $\delta_{\epsilon_+}\Psi_M=0$ is
\begin{equation}
\epsilon_+=e^{\left(\frac{\chi}{k_1}\,\Gamma_{\hat{7}}\Gamma_{\hat{8}}\right)}\,\epsilon_{0,+}\hspace{50pt}\text{with}\hspace{5pt}(1+\Gamma _{\hat{6}}\Gamma _{\hat{7}} \Gamma _{\hat{8}} \Gamma _{\hat{9}})\,\epsilon_{0,+}=0\,,
\end{equation}
where $\epsilon_{0,+}$ is a Majorana-Weyl spinor with constant entries. 


\paragraph{Gravitino variation for $\epsilon_-$}
We perform now the same analysis for $\epsilon_-$. The supersymmetry variations for the gravitino along the $M=i$ directions are 
\eq{
\arraycolsep2pt
\begin{array}{lcl}
\delta_{\epsilon_-}\Psi_r&=&\partial_r\epsilon_-\,,
\\[4pt]
\delta_{\epsilon_-}\Psi_\theta&=&\partial_\theta\epsilon_--\frac{1}{4 r^2 h(r)}\Big(2 N\,\Gamma _{\hat{8}} \Gamma _{\hat{9}} +r^2\left(\Gamma _{\hat{8}} \Gamma _{\hat{9}}-\Gamma _{\hat{6}} \Gamma _{\hat{7}} \right)\Big)\epsilon_-\,,  
\\
\delta_{\epsilon_-}\Psi_\xi&=&\partial_\xi\epsilon_--
\frac{1}{4k_2r^2 h(r)}\Big( \left(2 k_1k_2k_3+r^2\right) \Gamma _7\left( \sin \theta\,\Gamma _9+\cos\theta\,\Gamma _8 \right)
\\
&&\hspace{160pt}-r^2\Gamma _6  \left(\cos\theta\,\Gamma _9 -\sin\theta\,\Gamma _8 \right)\Big)\epsilon_-\,,
\\
\delta_{\epsilon_-}\Psi_\chi&=&\partial_\chi\epsilon_--\frac{1}{2 k_1h(r)}\left(\Gamma _{\hat{7}} \Gamma_{\hat{8}}+\Gamma _{\hat{6}} \Gamma _{\hat{9}}\right)\epsilon_- \,,
\end{array} 
}
and applying them to a spinor $\epsilon_-$ satisfying \eqref{eq:projection-condition-orbifold} they reduce to
\eq{
\arraycolsep2pt
\begin{array}{lcl@{\hspace{30pt}}lcl}
\delta_{\epsilon_-}\Psi_r&=&\partial_r\epsilon_-\,,
&
\delta_{\epsilon_-}\Psi_\xi&=&\bigl(\partial_\xi-\frac{1}{2 k_2}\sin\theta\,\Gamma _{\hat{7}} \Gamma _{\hat{9}}-\frac{1}{2 k_2}\cos\theta\,\Gamma _{\hat{7}} \Gamma _{\hat{8}}\bigr)\epsilon_-\,,
\\[8pt]
\delta_{\epsilon_-}\Psi_\theta&=&\bigl(\partial_\theta+\frac{1}{2}\Gamma _{\hat{8}} \Gamma _{\hat{9}}\bigr)\epsilon_-\,,
&
\delta_{\epsilon_-}\Psi_\chi&=&\partial_\chi\epsilon_-\,.
\end{array} 
}
The general solution to the supersymmetry equations is
\begin{equation}
\epsilon_-=e^{-\frac{\theta}{2}\Gamma_{\hat{8}}\Gamma_{\hat{9}}}e^{\frac{\xi}{2 k_2}\Gamma_{\hat{7}}\Gamma_{\hat{8}}}\epsilon_{0,-}\hspace{50pt}\text{with}\hspace{5pt}(1-\Gamma _{\hat{6}}\Gamma _{\hat{7}} \Gamma _{\hat{8}} \Gamma _{\hat{9}})\,\epsilon_{0,-}=0 \,,
\end{equation}
where $\epsilon_{0,-}$ is again a Majorana-Weyl spinor with constant entries.


\subsubsection*{T-dual configuration along $\chi$}
Next, we analyze the amount of supersymmetry preserved by the configuration obtained after performing a T-duality transformation along the direction $\chi$ to the NS5-orbifold, described by the fields \eqref{eq:NS5dual-chi}. We will find that part of the supersymmetry of the original background is broken by the T-duality transformation.


\paragraph{Dilatino variation}
The dilatino variation for the present background is now
\begin{equation}
\delta_\epsilon\lambda=-\frac{1}{2 r \sqrt{h(r)}}\,\Gamma_{\hat{6}}\, (1\pm\Gamma _{\hat{6}}\Gamma _{\hat{7}} \Gamma _{\hat{8}} \Gamma _{\hat{9}})\,\epsilon,
\end{equation}
which is again solved by a doublet of spinors satisfying
\begin{equation}\label{eq:projection-condition-orbifold-tdual}
(1\pm\Gamma _{\hat{6}}\Gamma _{\hat{7}} \Gamma _{\hat{8}} \Gamma _{\hat{9}})\,\epsilon_\pm=0\,.
\end{equation}
Note that in this case this condition is independent of the sign of $k_1k_2k_3$. As in the case before, this condition projects out half of the components of each Majorana-Weyl spinor.


\paragraph{Gravitino variation for $\epsilon_+$}
Although the dilatino variations do not depend on the sign of $k_1k_2k_3$, the gravitino variations do depend on it. For simplicity, we will only discuss the case where $k_1 k_2 k_3>0$. The case where $k_1 k_2 k_3<0$ can be discussed analogously, and the results are interchanged between the two components of the doublet. With the mentioned sign assumption, the variations $\delta_{\epsilon_+}\Psi_i$ for a spinor $\epsilon_+$ satisfying \eqref{eq:projection-condition-orbifold-tdual} are
\eq{
\arraycolsep2pt
\begin{array}{lcl}
\delta_{\epsilon_+}\Psi_r&=&\partial_r\epsilon_+\,,
\\[4pt]
\delta_{\epsilon_+}\Psi_\theta&=&\left(\partial_\theta+\frac{1}{2h(r)}\Gamma_{\hat{6}}\Gamma_{\hat{7}}\right)\epsilon_+ \,,
\\
\delta_{\epsilon_+}\Psi_\xi&=&\left(\partial_\xi+\frac{k_1k_2k_3}{2k_2r^2h(r)^2}\cos\theta\Gamma_{\hat{7}}\Gamma_{\hat{8}}+\frac{1}{2k_2h(r)}(\sin\theta\Gamma_{\hat{6}}\Gamma_{\hat{8}}+\cos\theta\Gamma_{\hat{7}}\Gamma_{\hat{8}})\right)\epsilon_+\,,
\\
\delta_{\epsilon_+}\Psi_\chi&=&\left(\partial_\chi-\frac{k_1^2k_2k_3}{r^4h(r)^2}\Gamma_{\hat{7}}\Gamma_{\hat{8}}\right)\epsilon_+\,.
\end{array} 
}
The first equation, $\delta_{\epsilon_+}\Psi_r=0$, is solved by spinors which are constant along the direction $r$. However, assuming this condition, the other equations $\delta_{\epsilon_+}\Psi_i=0$ cannot be solved. Therefore, for the present configuration there is no spinor $\epsilon_+$ satisfying the condition \eqref{eq:projection-condition-orbifold-tdual} with plus sign.


\paragraph{Gravitino variation for $\epsilon_-$}
For the case of $\epsilon_-$, the variations $\delta_{\epsilon_-}\Psi_i$ for a spinor $\epsilon_-$ satisfying \eqref{eq:projection-condition-orbifold-tdual} are
\eq{
\arraycolsep2pt
\begin{array}{lcl}
\delta_{\epsilon_-}\Psi_r &=&\partial_r\epsilon_-\,,
\\
\delta_{\epsilon_-}\Psi_\xi&=&\left(\partial_\xi-\frac{1}{2 k_2}\sin\theta\,\Gamma _{\hat{7}} \Gamma _{\hat{9}}-\frac{1}{2 k_2}\cos\theta\,\Gamma _{\hat{7}} \Gamma _{\hat{8}}\right)\epsilon_-\,,
\\
\delta_{\epsilon_-}\Psi_\theta&=&\left(\partial_\theta+\frac{1}{2}\Gamma _{\hat{8}} \Gamma _{\hat{9}}\right)\epsilon_-\,,
\\[4pt]
\delta_{\epsilon_-}\Psi_\chi&=&\partial_\chi\epsilon_-\,,
\end{array} 
}
which are the same as in the original NS5-orbifold. Therefore, the equations $\delta_{\epsilon_-}\Psi_M=0$ are solved again by
\begin{equation}\label{eq:epsilon-minus-solution-tdual}
\epsilon_-=e^{-\frac{\theta}{2}\Gamma_{\hat{8}}\Gamma_{\hat{9}}}e^{\frac{\xi}{2 k_2}\Gamma_{\hat{7}}\Gamma_{\hat{8}}}\epsilon_{0,-}\hspace{50pt}\text{with}\hspace{5pt}(1+\Gamma _{\hat{6}}\Gamma _{\hat{7}} \Gamma _{\hat{8}} \Gamma _{\hat{9}})\,\epsilon_{0,-}=0\,.
\end{equation}
This solution corresponds to the case where $k_1k_2k_3>0$. In the case where $k_1k_2k_3<0$, the $\epsilon_-$ component has no solution whereas the $\epsilon_+$ has a solution of the form \eqref{eq:epsilon-minus-solution-tdual}. In both cases, only half of the Killing spinors of the original background are present after the T-duality transformation. The configuration is then 1/4-supersymmetric.


\clearpage 
\bibliographystyle{utphys}
\bibliography{references}


\end{document}